\documentclass{article}
\usepackage{epsfig}

\begin{document}

\begin{flushright}
\normalsize
Freiburg-THEP 00/1\\
March 2000\vspace{0.8cm}\\
\end{flushright}
\begin{center}
{\Large\bf Regular and black-hole solutions of the Einstein-Yang-Mills-Higgs
equations;\\ \vspace{0.2cm} the case of nonminimal coupling}
\vspace{1.2cm}\\
J.J.\ van der Bij and Eugen Radu
\vspace{0.4cm}\\
\it Albert-Ludwigs-Universit\"at \\
\it Fakult\"at f\"ur Physik\\
\it Freiburg Germany\vspace{1.2cm}\\
\end{center}

\begin{abstract}

Regular  and black-hole  solutions  of the  spontaneously  broken Einstein-Yang-Mills-Higgs  
theory with nonminimal  coupling to gravity are shown to exist.  The main  characteristics 
of the solutions are presented and  differences  with respect to the  minimally  coupled 
case are studied. Since negative energy  densities are found to be possible, 
traversable  wormhole  solutions might exist.  We prove that they are absent.
\\
\\
\\
\\
\\
\\
\\
\\
\\
\\
\\
\end{abstract}
\newpage
\section{Introduction}

In the light of the  no-hair  \cite{1, 2} and  no-go  theorems  \cite{3,4} for the  
classical glueball  solutions with or without  gravity, the discovery of both 
smooth and black-hole solutions of the self-gravitating  non-Abelian gauge theories
was a big surprise (for a review see \cite{5}). Among such  solutions a  physically
interesting  case is that of a spontaneous  broken EYMH theory examined in \cite{6},
where both regular and black-hole  solutions, i.e. gravitating
sphalerons and sphaleron black-holes have been found. The stability analysis
of this system has shown that the solutions are unstable
\cite{7, 8}.

Because of the physical importance of these objects, it is worthwhile to study
generalizations of the couplings of the flat-space Lagangian to gravity.
 One of the simplest and best motivated extensions is the inclusion
 of an explicit  coupling  between the scalar field $\Phi$  and
the curvature of the spacetime $\mathcal{R}$ of the form $\xi\Phi^{\dagger}\Phi\mathcal{R}$,
 where $\xi$ is a dimensionless  coupling  constant.
There are many reasons to believe that a nonminimal coupling term appears.  A
nonminimal coupling is generated by quantum corrections even  if it is absent
in the classical action and is required in order to renormalize the theory
\cite{9}.

 In many  physical  situations,  inclusion  of a $\xi\ne0$ term  leads to new
interesting  physical  effects even at the classical level.
 Examples are
the Bronnikov-Melnikov-Bacharova-Bekenstein conformal scalar hair \cite{12, 13}, 
the inflationary scenario with a nonminimally coupled "inflaton" field
\cite{14, 15},  and boson star solutions \cite{16}. For a review of the
present situation see \cite{faraoni}. Two cases occur most frequently in the
literature: "minimal coupling" ($\xi$=0)  and "conformal coupling"
($\xi$=1/6). The conformal  invariance  dictates $\xi$=1/6 for a massless
scalar field \cite{9}, while  Nambu-Goldstone  bosons have a minimal coupling
$\xi$=0 \cite{17}.  However, there is no  preferential  value of $\xi$ for a 
Higgs field in a unified  gauge theory of electroweak   interactions.

In this paper we study numerically  regular and black-hole  
solutions of the coupled  EYMH field equations with a nonminimal coupling to 
gravity, extending the results of ref.\cite{6} to this case. 
Ref. \cite{6} presented strong numerical arguments for the existence of both regular 
and black-hole solutions in a minimally coupled EYMH theory. For each fixed value of 
the Higgs vacuum expectation value $v$, solutions have been found, that 
can be indexed by the number of nodes $k$ of the Yang-Mills potential function. For each
$k$ there are two branches of solution, depending on the behavior of $v\to 0$. 
For example, the so-called quasi $k=0$ branch of solutions approaches the
Schwarzschild solution as $v\to 0$; whereas the regular $k=1$ branch approaches
the first colored black-hole solution of the  Einstein-Yang-Mills system
\cite{bizon} in the same limit.  The two branches of solutions converge for
some values of the  theory parameters \cite{7}. 

Although most of the 
phenomena discussed in \cite{6} for the $\xi$=0 case repeat themselves in the 
general case, 
there are some important differences. For a nonminimal coupling, the time 
component of $T_{\mu\nu}$, which in Einstein's  gravity would correspond to the local 
energy density, may be non-positive.  Indeed,  as we shall see  later  on, 
there  are regions in space,  where this  quantity is negative.  The reason is that,
as a result of the  nonminimal coupling with gravity, there are contributions to $T_{\mu\nu}$ 
from the gravitational  field itself .  However, the local energy densities 
do not directly determine the sign of the asymptotic  ADM mass,  which is found 
to be positive. 

Also, the parameter range of
the solutions found in ref.\cite{6} remains no longer valid and a new range has to be 
found for every choice of $\xi$. The existence of a nonminimal coupling between the Higgs
 field and the gravitational field implies a decrease of the maximal allowed vacuum 
 expectation 
value of the Higgs field. 

The  paper is structured as follows:  in section II we present the
general framework and an analysis of the field equations, while in section III
we adress the problem of the numerical  construction  of  solutions.  In section IV
the possibility of the existence of Lorentzian wormholes is considered with a negative
result.  We conclude with section V where the results are compiled.

\section{GENERAL FRAMEWORK AND BASIC EQUATIONS}

Our study of the EYMH system is based upon the action
\begin{equation} \label{lag0}
S=\int d^{4}x\sqrt{-g}[\frac{\mathcal{R}}{16\pi G}-
\frac{1}{4\pi}((D_{\mu}\Phi)^{\dagger}(D^{\mu}\Phi)+
V(\Phi)+\xi\Phi^{\dagger}\Phi\mathcal{R})-\frac{1}{4\pi}\frac{1}{4}\mid F\mid^{2}]. 
\end{equation}
Here  $G$ is  the  gravitational constant,  $D_{\mu}$ is  the  usual  gauge-covariant
derivative expressed in the anti-hermitian basis of SU(2) ($\tau_{a}=-i\sigma_{a}/2$)
\begin{equation}
D_{\mu}=\partial_{\mu}+g\tau\cdot A_{\mu},
\end{equation}
$g$ is the gauge coupling constant. Following \cite{6}, we assume that $\Phi$
possesses  only one degree of freedom
\begin{equation}
\Phi=\frac{1}{\sqrt{2}}{0 \choose \phi(x)},
\end{equation}
with $\phi$ real and time independent, and with the Higgs potential
\begin{equation}
V(\phi)=\frac{\lambda}{4}(\phi^{2}-v^{2})^{2},
\end{equation}
where $v$ denotes the vacuum expectation value of $\Phi$. 
The action (\ref{lag0}) becomes
\begin{equation} \label{lag}
S=\int d^{4}x\sqrt{-g}[\frac{\mathcal{R}}{16\pi G}-\frac{1}{4\pi}\frac{1}{2}
((\partial_{\mu}\phi)(\partial^{\mu}\phi)+(\frac{g\phi}{2})^{2}
A_{\mu}A^{\mu}+
V(\phi)+\xi\phi^{2}\mathcal{R})-\frac{1}{4\pi}\frac{1}{4}\mid F\mid^{2}]. 
\end{equation}
There are considerable  modifications in the Einstein equations due to the new
energy-momentum tensor:
\begin{eqnarray}
8\pi T_{\mu\nu} =&& 8\pi T_{\mu\nu}^{(minimal)}+2\xi(G_{\mu\nu}\phi^{2}
+g_{\mu\nu}\nabla_{\gamma}\nabla^{\gamma}\phi^{2}
-\phi^{2}_{,\mu ;\nu})
\\
8\pi T_{\mu\nu}^{(minimal)} = &&2F_{\mu\gamma}F_{\nu}^{{ }{ }\gamma}-
\frac{1}{2}g_{\mu\nu}\mid F\mid^{2}+2(\frac{g\phi}{2})^{2}
A_{\mu}A_{\nu}
-g_{\mu\nu}(\frac{g\phi}{2})^{2}A_{\gamma}A^{\gamma}
\nonumber\\
&&+2(\partial_{\mu}\phi)(\partial_{\nu}\phi)-
g_{\mu\nu}((\partial_{\gamma}\phi)(\partial^{\gamma}\phi)+2V(\phi)).
\end{eqnarray}
where $G_{\mu\nu}$ is the Einstein tensor.
As we assume spherical symmetry it is convenient to use the usual metric form:
\begin{equation} \label{metric}
ds^{2}=R^{2}(r)dr^{2}+r^2(d\theta^{2}+sin^{2}\theta d\varphi^{2})-
\frac{dt^{2}}{T^{2}(r)}
\end{equation}
where $R(r)=(1-2m(r)/r)^{-1/2}$ and $m(r)$ may be interpreted  as the total  mass-energy  
within the radius $r$.  To describe the black-hole solutions we define  $\delta=-ln(R/T)$; thus:
\begin{equation}
ds^{2}=(1-\frac{2m(r)}{r})^{-1}dr^{2}+r^2(d\theta^{2}+sin^{2}\theta d\varphi^{2})-
(1-\frac{2m(r)}{r})e^{-2\delta(r)}dt^{2}.
\end{equation}
The event horizon is at $r=r_{h}$  where  $1/R^{2}(r_{h})=0$. In case there are several such 
zeroes,
the horizon corresponds to the outer one. Regularity at the origin is satisfied 
when T(0)$<\infty$ and  $R'(0)=T'(0)=0$,  while regularity  at the  event  horizon  
r=$r_{h}$ requires $\delta$($r_{h})<\infty$. In this paper we deal with nonextremal 
black-holes only, i.e. near the event horizon
\begin{equation}
1-2m(r)/r\sim r-r_{h}.
\end{equation}
A suitable rescaling of the time coordinate $t$ implies:
\begin{eqnarray}
R(0)=1,
\quad T(0)=1,
\quad m(r_{h})=r_{h}/2,
\quad \delta(r_{h})=0.
\end{eqnarray}

For the  Yang-Mills field,  it is  convenient  to use the  ansatz
discussed in \cite{6}; thus a suitable parametrization of the Yang-Mills connection
is:
\begin{equation}
A=\frac{1}{g}(1+\omega)[-\widehat\tau_{\varphi}d\theta+\widehat\tau_{\theta}  \sin \theta  d\varphi].
\end{equation}
The $\widehat\tau_{i}$ are appropriately normalised spherical generators of the SU(2) group in 
the notation of ref. \cite{6}, $e.g.$ $\widehat\tau_{r}=\widehat r\cdot\tau$, 
$[\tau_{a},\tau_{b}]=\epsilon_{abc}\tau_{c}$,  
while $\phi(r)$ is the Higgs field.

Expressing  the curvature  scalar $\mathcal{R}$ in terms of the metric  function  $R(r)$ and
$T(r)$, we obtain the following  expression of the reduced  action of our static
spherically symmetric system:
\begin{eqnarray}
S=\int drdt 
[\frac{1}{2G}\frac{1}{T}(R-\frac{1}{R}+2r\frac{R'}{R})-
\frac{1}{2}(\frac{(\phi ')^{2}r^{2}}{RT}+
\frac{\phi^{2}(1+\omega)^{2}}{2}\frac{R}{T})
\nonumber\\
-V(\phi)r^{2}\frac{R}{T}-
\frac{1}{g^{2}}(\frac{(\omega ')^{2}}{RT}+
\frac{(1-\omega^{2})^{2}}{2r^{2}}\frac{R}{T})-
\xi\frac{\phi^{2}}{RT}(R^{2}-1+2r\frac{R'}{R})+
2\xi\phi\phi '\frac{r^{2}T'}{RT^{2}}]
\end{eqnarray}
for a regular spacetime, while a suitable form of the reduced action for a black-hole 
spacetime is
\begin{eqnarray}
S=\int drdt e^{-\delta}
[\frac{m'(1-2\xi G\phi^{2})}{G}-
\frac{1}{2}((\phi ')^{2}r^{2}(1-\frac{2m}{r})+
\frac{\phi^{2}(1+\omega)^{2}}{2})
\nonumber\\
-V(\phi)r^{2}-
\frac{1}{g^{2}}((\omega ')^{2}(1-\frac{2m}{r})+
\frac{(1-\omega^{2})^{2}}{2r^{2}})+
2\xi\phi\phi '(r^{2}\delta '(1-\frac{2m}{r})+m'r-m)]
\end{eqnarray}
where the prime denotes derivative with respect to r.

A  usual  rescaling  \cite{18}
\begin{eqnarray} \label{rescaling}
r\to rg/\sqrt{G},
\quad 
\phi\to\phi/\sqrt{G}
\end{eqnarray}  
 reveals  that  the  solutions  depend
essentially  on two  dimensionless  parameters $\alpha$ and $\beta$, 
 expressible  through the mass
ratios
\begin{eqnarray} 
\alpha=\frac{M_{W}}{v M_{Pl}};
\quad
 \beta=\frac{M_{H}}{M_{W}}
\end{eqnarray}
with $M_W=gv$, $M_H=\sqrt{\lambda}v$ and $M_{Pl}=\frac{1}{\sqrt{G}}$ . 
$V(\phi)=\frac{\beta^{2}}{4}(\phi^{2}-\alpha^{2})^{2}$ is
 the standard double well field potential.
The field equations imply the relations
\begin{equation} \label{weq}
\omega''=\omega'(\frac{R}{R'}+\frac{T}{T'})+
\frac{\omega(\omega^{2}-1)}{r^{2}}R^{2}+
\frac{\phi^{2} (\omega+1)}{4}R^{2}
\end{equation}
for the gauge field, and
\begin{equation} \label{higseq}
\phi''=\phi'(\frac{R}{R'}+\frac{T}{T'}-\frac{2}{r})+
R^{2}[\frac{\phi (1+\omega)^{2}}{2r^{2}}+\frac{dV}{d\phi}
+\xi \mathcal{R}\phi]
\end{equation}
for the Higgs field, where 
\begin{eqnarray} \label{R}
\mathcal{R}=\frac{1}{\frac{1}{2}+(6\xi-1)\xi\phi^{2}}((1-6\xi)(\phi ')^{2}
(1-\frac{2m}{r})+4V(\phi)+\frac{\phi^{2}(1+\omega)^{2}}{2r^{2}}
\nonumber\\
-6\xi\phi\frac{dV(\phi)}{d\phi}-
3\xi\phi^{2}\frac{(1+\omega)^{2}}{r^{2}})
\end{eqnarray}
is the spacetime curvature.
 The $(rr)$ and $(tt)$ Einstein equation are
\begin{eqnarray} \label{meq}
(1-2\xi\phi^{2})m'=
(1-\frac{2m}{r})(\frac{(\phi')^{2}r^{2}}{2}+(\omega')^{2}
-2\xi\phi\phi 'r^{2}\frac{T'}{T}
-2\xi r^{2}(\phi ')^2)
+Vr^{2}+
\nonumber\\
\frac{\phi^{2}(1+\omega)^{2}}{4}
+\frac{(1-\omega^{2})^{2}}{2r^{2}}-
2\xi r^{2}(\phi\frac{dV}{d\phi}+
\frac{\phi^{2}(1+\omega)^{2}}{2r^{2}}+
\xi \mathcal{R}\phi^{2}),
\end{eqnarray}
\begin{eqnarray} \label{Teq}
(\xi\phi\phi 'r-(\frac{1}{2}-\xi\phi^{2}))2r\frac{T'}{T}=
(\frac{1}{2}-\xi\phi^{2})\frac{2m}{r}\frac{1}{1-\frac{2m}{r}}+
4\xi\phi\phi 'r+
(\omega ')^{2}+
\frac{r^{2}(\phi)'^{2}}{2}
\nonumber\\
-\frac{1}{(1-\frac{2m}{r})}(Vr^{2}+
\frac{\phi^{2}(1+\omega)^{2}}{4}
+\frac{(1-\omega^{2})^{2}}{2r^{2}}).
\end{eqnarray}
For the black-hole  solutions  we replace the  auxiliary  $T'$ equation  with an
equation for $\delta'$
\begin{eqnarray} \label{deltaeq}
r\delta '(-1+2\xi\phi^{2})=a_{1}+\frac{a_{2}}{1-\frac{2m}{r}}-
\frac{\frac{1}{2}+\xi\phi\phi 'r-\xi\phi^{2}}
{-\frac{1}{2}+\xi\phi\phi 'r+\xi\phi^{2}}
(a_{3}+\frac{a_{4}}{1-\frac{2m}{r}})
\end{eqnarray}
where
\begin{eqnarray}
a_{1} =&& (\omega ')^{2}+\frac{(\phi ')^{2}r^{2}}{2}-2\xi r^{2}(\phi ')^{2},\\
a_{2} =&& -(\frac{1}{2}-\xi\phi^{2})\frac{2m}{r}
+Vr^{2}+\frac{\phi^{2}(1+\omega)^{2}}{4}
\nonumber\\
&&+\frac{(1-\omega^{2})^{2}}{2r^{2}}-
2\xi r^{2}(\phi\frac{dV}{d\phi}+
\frac{\phi^{2}(1+\omega)^{2}}{2r^{2}}+
\xi \mathcal{R}\phi^{2}),\\
a_{3} =&& 4\xi\phi\phi 'r+
(\omega ')^{2}+
\frac{r^{2}(\phi)'^{2}}{2},\\
a_{4} =&& (\frac{1}{2}-\xi\phi^{2})\frac{2m}{r}-
(Vr^{2}+
\frac{\phi^{2}(1+\omega)^{2}}{4}
+\frac{(1-\omega^{2})^{2}}{2r^{2}}).
\end{eqnarray}
Following the analysis in \cite{6}, we can already predict the 
boundary conditions and some general features of the finite energy solutions.
 Since the Yang-Mills equations are unaffected
 by the presence of the $\xi\Phi^{2}\mathcal{R}$ term in (\ref{lag}), the  analysis 
  presented  by Greene, Mathur and O'Neill \cite{6} 
  remains
valid:  $\omega=-1$ is the  only
acceptable  value and $\omega\leq1$ is  required  for finite  energy  solutions.  If we
assume that $\phi$  is $O(\alpha)$ in the region $r\geq1$, we obtain:
\begin{eqnarray}
r\leq O(1/\alpha) : -1<\omega<\frac{1}{2}(1+\sqrt{1-(\alpha r)^{2}})
\nonumber\\
r\geq O(1/\alpha) : \omega '<0, \omega ''>0
\end{eqnarray}
These constraints  are valid for both regular and black-hole solutions.  The
Higgs equation can be written in the form
\begin{eqnarray}\label{fi2}
\frac{1}{2}\frac{d}{dr}(\frac{r^{2}(\phi^{2})'}{RT})=
\frac{1}{\frac{1}{2}+(6\xi-1)\xi\phi^{2}}[\frac{1}{2}\frac{(\phi ')^{2}
r^{2}}{RT}+\frac{Rr^2}{T}(\frac{\phi}{2}\frac{dV}{d\phi}+
\frac{(1+\omega)^{2}\phi^{2}}{4r^{2}})
\nonumber\\
+2\xi\phi^{2}r^{2}\frac{R}{T}(2V-\frac{\phi}{2}\frac{dV}{d\phi})].
\end{eqnarray}
The obvious requirement for finite energy solution is
\begin{eqnarray} \label{rel1}
\phi\frac{dV}{d\phi}(1-2\xi\phi^{2})<0
\end{eqnarray}
which implies that $\phi$ is restricted to lie between the minima of the potential,
$-\alpha\leq\phi\leq\alpha$.  Relation (\ref{rel1}) provides also an upper bound on 
the range of  $\xi$ :
\begin{equation} \label{csi}
\xi<\frac{1}{2\alpha^{2}}
\end{equation}
It is worth noting, that  an investigation \cite{19} of  classical  stability
of a scalar field in a curved spacetime  with a general coupling to gravity
found, that the Higgs fields in the  standard 
 model must have  $\xi\leq0$ or $\xi \geq 1/6 $.

>From the relation (\ref{lag}) we can see that an effective gravitational
constant  is given by
\begin{equation}
G_{eff}=\frac{G}{1-2\xi\phi^{2}}.
\end{equation}

Thus the  condition (\ref{csi}) implies  positivity  of the  effective  gravitational
constant.  It is not  possible  to obtain an explicit lower bound for $\xi$ 
for a given value of $\alpha$.        
The field equations imply that $\pm\alpha$ are the only  allowed  values of $\phi$ 
as $r\to\infty$.
We focus  here on  solutions  with 
  $\phi(\infty)=\alpha$ without loss of generality. 
The vacuum values $\omega(\infty)=-1$ and 
  $\phi(\infty)=\alpha$    
are shared both by  black-holes and regular solutions. 
The analysis of the field equations as $r\to\infty$ gives
\begin{eqnarray} \label{masimpt}
m(r)\sim && M +\frac{1}{1-2\xi\alpha^{2}}(2a\sqrt{2}\xi\alpha^{2}\beta c r^{2}
e^{-\sqrt{2}\alpha\beta cr}
\nonumber\\
&&-\frac{\alpha\beta}{2\sqrt{2}c}((c^{2}+1)(1-4\xi)+\frac{48\xi^{3}\alpha^2}
{\frac{1}{2}+(6\xi-1)\xi\alpha^{2}}))
a^{2}r^{2}e^{-2\sqrt{2}\alpha\beta cr},
\label{Tasimp}\\
lnT(r)\sim &&ln(T_{0})+\frac{M}{r},
\label{deltasimpt}\\
\delta (r)\sim &&-\delta_{0}
-2\sqrt{2}\frac{\xi\alpha^{2}\beta c}{1-2\xi\alpha^{2}}ar
e^{-\sqrt{2}\alpha\beta cr}
\nonumber\\
&&+\frac{\alpha\beta}{2\sqrt{2}c}((c^{2}+1)(1-4\xi)+\frac{48\xi^{3}\alpha^2}
{\frac{1}{2}+(6\xi-1)\xi\alpha^{2}})
\frac{a^{2}re^{-2\sqrt{2}\alpha\beta cr}}{1-2\xi\alpha^{2}},
\label{wasimpt}\\
\omega (r)\sim &&-1+be^{-\frac{\alpha r}{2}},
\label{fiasimpt}\\ 
\phi (r)\sim &&\alpha+ae^{-\sqrt{2}\alpha\beta cr},
\end{eqnarray}
where $c=\sqrt{\frac{1-2\xi\alpha^{2}}{1+2\xi\alpha^{2}(6\xi-1)}}$; 
$M, b, a$ are constants; $b>0, a<0$.

Relation (\ref{masimpt}) implies an  asymptotic  violation  of the weak energy condition
(WEC) for negative values of $\xi$, since $m'(r)<0$. There exist also other 
classical  field  theories  that  violate the  WEC. 
Examples are theories  
 containing  $\mathcal{R}+\mathcal{R}^{2}$  
terms in the action\cite{20} ,  an  antisymmetric  3-form
axion  field  coupled  to  scalar  fields  \cite{21}, the Brans-Dicke scalar-tensor 
theory \cite{22},  and Einstein-dilaton theory with
curvature-squared terms  of Gauss-Bonnet type \cite{23}.

Since  $\mu (\xi)=\alpha \beta \sqrt{\frac{1-2\xi\alpha^{2}}{1+2\xi\alpha^{2}(6\xi-1)}}$ 
corresponds to the mass of the Higgs field at infinity, 
the following relation holds
\begin{equation}
\mu (\xi)\leq \mu (0)=\alpha\beta.
\end{equation}
Thus any nonminimal coupling decreases the asymptotic value of the Higgs field
 mass. 
To get further insight into the meaning of a large value of $\xi$ it is 
worthwhile to consider the rescaling
$r\to r/\sqrt{(-\xi)}$, 
$\phi\to\phi/\sqrt{(-\xi)}$. 
For $\xi\to-\infty$ we find that $\alpha\to0$ necessarily. Thus, for a large 
negative $\xi$, we expect a decrease 
of the maximal allowed value of the parameter $\alpha$. Furthermore, 
there is an effective  decoupling of the Yang-Mills  and gravitational fields
 and the
effective  coupling  of the  Higgs  field to matter  becomes  of  gravitational
strength.
In the limit of  infinite negative $\xi$, the following field equations are
 obtained 
\begin{eqnarray}
&&(\frac{1}{2}+\phi^{2})2r\frac{R'}{R}=(\frac{1}{2}+\phi^{2})(1-R^{2})+
2\phi\phi 'r^{2}\frac{T'}{T},
\\
&&(\phi\phi 'r+\frac{1}{2}+\phi^{2})2r\frac{T'}{T}=
(\frac{1}{2}+\phi^{2})(1-R^{2})+4\phi\phi 'r,
\\
&&\omega''=\omega'(\frac{R'}{R}+\frac{T'}{T})+\frac{\omega (\omega^{2}-1)R^{2}}{r^{2}}+
\frac{\phi^{2}(1+\omega)^{2}}{4}R^{2},
\\
&&\phi''=\phi'(\frac{R'}{R}+\frac{T'}{T}-\frac{2}{r})-
\frac{(\phi ')^{2}}{\phi}.
\end{eqnarray}
By using a usual power series expansion near the origin or the event horizon it
can be proven 
that there are no initial conditions consistent with the requirement of energy 
finiteness.
However a simpler proof is to observe that equation (41) implies the relation
$(\phi^{2})'=const.\frac{RT}{r^{2}}$, which is also consistent with the general 
equation (\ref{fi2}). There are no  nonsingular solution of this
 equation consistent with the requirement of metric regularity at the origin 
 or with a regular event horizon. Thus we conclude that nontrivial solutions 
are absent in the case of an infinite negative $\xi$.
In practice, it becomes increasingly difficult to solve 
 the field equations for large negative values of $\xi$, with a fast 
convergence  to the asymptotic
 values $\omega(\infty)$, $\phi(\infty)$.

A  particularly  interesting  case of the general theory is obtained for
 $G\to\infty$, i.e. in the 
absence of the Einstein term in the action (\ref{lag}), 
corresponding to the spontaneous 
symmetry breaking theory of gravity, with the standard Higgs field as the origin
of the Plank mass (for the remainder of this paragraph we do not consider
 the rescaling 
(\ref{rescaling})).

There has  recently been  an  increased  interest  in induced  gravity
in the standard  model, that might help to solve  some  problems  of 
particle physics and cosmology. 
Typical problems are  the necessity of the Higgs mass to be
order  of the theory cut-off \cite{24}, the missing mass  problem,  Mach's  
principle \cite{25},
and the inflationary  scenario \cite{26}. 
The existence of magnetic monopole solutions in an induced gravity YMH theory
has  also been discussed \cite{magnetic monopole1}.
 Unfortunately, it can be proven that,
in the absence of the Einstein term,  only the trivial case $\phi=\pm\alpha$
is   consistent  with the requirement of energy finiteness. 
  When we consider the  equation (\ref{higseq}) and use
the trace of Einstein equations to eliminate the term $\xi\phi\mathcal{R}$ we obtain the
general equation
\begin{equation}
\frac{1}{2}\nabla_{\mu}\nabla^{\mu}\phi^{2}=\frac{1}{1-6\xi}
(\frac{\mathcal{R}}{2G}+\phi\frac{dV}{d\phi}-4V).
\end{equation}
For our ansatz we obtain the relation
\begin{equation}
\frac{d}{dr}(\frac{r^{2}(\phi^{2})'}{RT})=\frac{Rr^{2}}{T}\frac{1}{1-6\xi}
(\frac{\mathcal{R}}{2G}+\phi\frac{dV}{d\phi}-4V).
\end{equation}
which should be satisfied for all $r$.  Since  clearly $4V-\phi\frac{dV}{d\phi}>0$
 for the considered potential, 
it follows that 
in the absence of an Einstein term in the original action only $\phi=\pm\alpha$ is 
consistent with the assumption of finite energy, both  for regular and
black-hole solutions.  However, for $\phi=\pm\alpha$
 we obtain the Bartnik-McKinnon solutions and their black-hole generalizations; 
 there are no spherically symmetric gravitating sphaleron or 
 sphaleron black-hole solutions.
 One can conjecture that similar to the boson star case, it may be
  possible to obtain nontrivial 
solutions by considering a time dependence of the matter field. 

\section{NUMERICAL SOLUTION}

Nontrivial solutions are not known  in closed form, and so a numerical method of 
solution is necessary.
\\
\subsection{REGULAR SOLUTIONS}

For regular solutions, finite $T_{tt}$ and regularity of the metric at $r=0$ give
 two possible sets of initial conditions
\begin{eqnarray} 
2m(r)=O(r^{3})
\\
lnT(r)=O(r^{2})
\\
{\omega (r) \choose \phi (r)}={-1+O(r^{2}) \choose \phi_{0}+O(r^{2})},
\end{eqnarray}
or 
\begin{eqnarray}
{\omega (r) \choose \phi (r)}={1+O(r^{2}) \choose O(r)}.
\end{eqnarray}
The general properties of the solutions are the same as for the minimally-coupled
EYMH  theory.  Solutions  are again  characterized  by $w(r)$  oscillations  in the
region $r>1$ and classified by the node number k which may be even or odd.  The
formal power series describing the above boundary conditions at $r=0$ is
\begin{eqnarray} 
&&2m(r)=ar^{3}+O(r^{5}),
\\
&&lnT(r)=\frac{1}{2}fr^{2}+O(r^{4}),
\\
&&\omega=-1+br^{2}+O(r^{4}),
\\
&&\phi=\phi_{0}+er^{2}+O(r^{4}),
\end{eqnarray}
with
\begin{eqnarray}
&&a=\frac{4b^{2}+\frac{2}{3}V_{0}}{1-2\xi\phi_{0}^{2}}
-\frac{4}{3}\frac{\xi\phi_{0}}{1-2\xi\phi_{0}^{2}}
\frac{(\frac{1}{2}-\xi\phi_{0}^{2})V_{0}'+4V_{0}\phi_{0}\xi}
{\frac{1}{2}+(6\xi-1)\xi\phi_{0}^{2}},
\\
&&f=-\frac{1}{2}\frac{4b^{2}+
\frac{2}{3}V_{0}}{1-2\xi\phi_{0}}
+\frac{2}{3}\frac{\xi\phi_{0}}{1-2\xi\phi_{0}^{2}}
\frac{(\frac{1}{2}-\xi\phi_{0}^{2})V_{0}'+4V_{0}\phi_{0}\xi}
{\frac{1}{2}+(6\xi-1)\xi\phi_{0}^{2}}
\nonumber\\
&&+\frac{8\xi\phi_{0}e-V_{0}+2b^{2}}{2\xi\phi_{0}^{2}-1},
\\
&&e=\frac{1}{6}V_{0}'
+\frac{\xi\phi_{0}}{\frac{1}{2}+(6\xi-1)\xi\phi_{0}^{2}}
(4V_{0}-6\xi\phi_{0}V_{0}'),
\end{eqnarray}
for  even-k  solutions  ($V_0, V'_0$   are the  potential  and its  derivative  with
respect to $\phi$ at $\phi=\phi_0$) and
\begin{eqnarray}
&&2m(r)=(4b^{2}+\frac{2}{3}V_{0}+e^{2}-4\xi e^{2})r^{3}+O(r^{5}),
\\
&&lnT(r)=-(2b^{2}-\frac{1}{3}V_{0}+2\xi e^{2})r^{2}+O(r^{4}),
\\
&&\omega =1-br^{2}+O(r^{4}),
\\
&&\phi =er+O(r^{3}),
\end{eqnarray}
for odd-k solutions.  The shooting parameters are $(\phi_0,b)$ and $(e,b)$ respectively.
  Using a standard ordinary  differential  equation solver, we
evaluate  the  initial  conditions  at  $r=10^{-3}$ for  global  tolerance  $10^{-12}$,
 adjusting  for fixed shooting parameters and  integrating  towards  $r\to\infty$.
  The  difficulty  of the
two-dimensional  shooting problem in the presence of two free parameters
 is increased by the presence of a nonminimal  coupling  between the Higgs field
and  gravity which leads to a slow convergence of the mass function $m(r)$.


We limit the discussion to results containing only one or two nodes. 
The results obtained for the $k=1$ and $k=2$ solutions retain 
the general characteristics of the minimally coupled case. In order
to define the terminology, which is somewhat confusing, we review here the
$\xi = 0$ case (\cite{5, 6, 18}). 

In the $k=1$ case there are two possible solutions that are
called the  quasi k=0 solution and the proper $k=1$ solution. The different
character of the two branches becomes apparent in the limit $\alpha
\rightarrow 0$. Remembering  that $\alpha = v \sqrt{G}$, there are two 
physically inequivalent ways for $\alpha$ to approach $0$. The first is
for the Newton constant to vanish, the second for the Higgs vacuum expectation
value to vanish. In the first case we have weakly coupled gravity and the
solution approaches the standard model sphaleron. 
The quasi-$k=$0 branch of solutions has this limit for $\alpha \rightarrow 0$.
For $\alpha \rightarrow 0$ the node that is present
actually moves to infinity in this branch, therefore the name quasi-$k=$0.  
The proper $k=1$ branch approaches  the first Bartnik-McKinnon solution
when $\alpha \rightarrow 0$.  It corresponds to taking $v \rightarrow 0$, but
always having gravity present.  For small enough $\alpha$, depending
on $\beta$,  both  solutions are present and different. For finite $\alpha$
the  behaviour is dependent on the value of $\beta$. For $\beta$ larger than a
critical value $\beta_{crit} \approx 0.12$ there is a first maximum value of
$\alpha$, where the proper $k=1$ branch disappears. For a larger value 
of $\alpha$ also the quasi-$k=0$ solution disappears. In the case $\beta$ smaller
than $\beta_{crit}$ the maximum value of $\alpha$ is the same for both
branches. Here the solutions merge; the shooting parameters approach each
other. The situation is clarified in $figure$ 1.

For the case of two nodes the situation is similar.  There are again two
branches, called the quasi$-k=1$ solution and the proper $k=2$ solution.
The quasi-$k=1$ solution approaches the first Bartnik-McKinnon solution
in the limit $\alpha \rightarrow 0$,  one of the nodes moving to infinity.
The proper $k=2$ solution approaches the second Bartnik-McKinnon solution.
The behaviour as a function of $\alpha$ and $\beta$ is qualitatively similar
to the one-node case ($figure$ 1). 

   To compare numerically the results with those found in
\cite{6} we  focused on solutions with $\beta^{2}=1/8$ (although similar
results have been obtained for other choices of $\beta$) and with $k=1$
and $k=2$ only. 
The results of the numerical  integration for  $\alpha$=0.1,  $\beta^{2}$=1/8 
and a range of  $\xi$ are  presented in $figure$ 2.

In this figure $\xi$  is relatively small (i.e.
$\xi\ll\frac{1}{2\alpha^{2}}$) . As a consequence the results  of \cite{6}
remain approximately valid.  The correction to the shooting parameters is very
small.
A general feature of the solutions is the small influence of the term $\xi\Phi^{2}\mathcal{R}$ 
on 
the value of the ADM mass. A negative value  of $\xi$  seems to decrease the asymptotic
 value of the
$m(r)$ function while a positive $\xi$ determines a higher ADM mass with respect to the 
minimally coupled case. The effect is particularly small for the quasi-$ k=0$
solution.  This is understandable as this solution corresponds to the flat
space sphaleron, for which gravity is a small effect.  For the considered range
of $\xi$, a nonminimal coupling term has a small effect on the shape  of the
$w(r)$ function. However, for suitable values of $\xi$ we have noticed a strong
influence  of this term on the  behavior of the Higgs field in the
intermediate region. Also, for k=2 solutions,  the initial value of the Higgs
field $\phi_{0}$ is strongly dependent on the value of $\xi$, with 
$\phi_{0}(\xi \neq 0)<\phi_{0}(\xi = 0)$ ($figure$ 2 c, d) . As we expected
from (\ref{masimpt}),  for  negative $\xi$ we obtain a violation  of the WEC  
beyond a certain limit of the radial coordinate, corresponding to a peak of
the mass   function $m(r)$. The height of the peak is proportional to the
absolute value of $\xi$.

In $figure$3 we study the solutions as a function of $\alpha$.
Significant changes occur for $\xi\to\frac{1}{2\alpha^{2}}$ and for large
negative values  of $\xi$. The  parameter range obtained in \cite{6} for the
two sheets of  solutions does not remain  valid; it is necessary to establish
a different value of $\alpha_{max}$ for every  choice of $\xi$. A general
feature for a nonminimal coupling  is the decrease of the maximal 
allowed value of the parameter $\alpha$ . 
For example for the proper $k=1$ branch, Greene, Mathur and O'Neill \cite{6}
 have found
$0<\alpha<0.599$; for $\xi=0.1$ we have obtained $0<\alpha<0.48$, while for $\xi=-2$ the 
limiting 
value of the parameter $\alpha$ is 0.51 ($figure$ 3a, b). The quasi-k=0 branch with $\xi=0$ has 
$0<\alpha<0.619$; for $\xi=-1$ we have found $\alpha_{max}=0.616$ while for $\xi=1/6$ we have 
$0<\alpha<0.525$ ($figure$ 3c, d). A minimally coupled even $k$ configuration has 
$0<\alpha<0.120$ (proper $k=2$ branch) 
or $0<\alpha<0.122$ (quasi-$k=1$ branch); in these cases, for $\xi=-1$ or $\xi=1/6$ we did 
not notice a significant deviation of the 
$\alpha_{max}$ value ($figure$ 3 e-h).
 Different limiting values occur for the shooting 
parameters $b, \phi_{0}$ and $e$ also.


\subsection{BLACK-HOLE SOLUTIONS}

Similar  results can be obtained for numerical  black-hole  solutions.  We use
the following expansion near the event horizon:
\begin{eqnarray} \label{m(rh)}
m(r)=\frac{r_{h}}{2}+m'(r_{h})(r-r_{h}),
\end{eqnarray}
\begin{eqnarray} \label{delta(rh)}
\delta(r)=0+\delta '(r_{h})(r-r_{h}),
\end{eqnarray}
\begin{eqnarray} \label{w(rh)}
\omega (r)=\omega (r_{h})+\omega '(r_{h})(r-r_{h}),
\end{eqnarray}
\begin{eqnarray} \label{fi(rh)}
\phi (r)=\phi (r_{h})+\phi '(r_{h})(r-r_{h}),
\end{eqnarray}
with
\begin{eqnarray}
&&m'(r_{h})=\frac{1}{2}+\frac{1}{2}\frac{a_{2}(r_{h})-a_{4}(r_{h})}
{1-2\xi\phi^{2} (r_{h})},
\\
&&\omega '(r_{h})=-\frac{r_{h}(1-2\xi\phi^{2} (r_{h}))}{a_{2}(r_{h})-a_{4}(r_{h})}
(\frac{\omega(r_{h})(\omega^{2}(r_{h})-1)}{r_{h}^{2}}
\nonumber\\
&&+\frac{\phi^{2} (r_{h})(1+\omega (r_{h}))}{4}),
\\
&&\phi '(r_{h})=\frac{1-2\xi\phi^{2}(r_{h})}{2\xi\phi(r_{h})r_{h}}
\frac{a_{2}(r_{h})+a_{4}(r_{h})}{a_{2}(r_{h})-a_{4}(r_{h})},
\\
&&\delta '(r_{h})=\frac{1}{r_{h}(1-2\xi\phi^{2} (r_{h}))}
(\frac{f '(r_{h})r_{h}}{1-2m'(r_{h})}\frac{1}
{-\frac{1}{2}+\xi\phi^{2}(r_{h})+\xi\phi(r_{h})\phi '(r_{h})r_{h}}
\nonumber\\
&&+\frac{(a_{1}(r_{h})+a_{3}(r_{h}))(\frac{1}{2}-\xi\phi^{2}(r_{h}))-
\xi\phi(r_{h})\phi '(r_{h})r_{h}(a_{1}(r_{h})-a_{3}(r_{h}))}
{-\frac{1}{2}+\xi\phi^{2}(r_{h})+\xi\phi(r_{h})\phi '(r_{h})r_{h}})
\end{eqnarray}
where
\begin{eqnarray}
f(r)=a_{2}(r)+a_{4}(r))(\frac{1}{2}-\xi\phi^{2})-\xi\phi\phi 'r
(a_{2}(r)-a_{4}(r)).
\end{eqnarray}
The new shooting parameters are $\omega (r_{h})$ and  $\phi(r_{h})$ (we have studied the 
case $r_{h}=1$ only).

The non-minimal gravitational coupling allows for a  not necessarily positive
field energy. Therefore  one loses one of the earlier tools for proving the no
hair theorems, which  already failed for the minimally coupled EYMH system. The
bypassing of the usual no-hair theorems in the considered system can be proven
by using the method of  \cite{7} for  the $\xi=0$ case.

Starting from the solutions (\ref{m(rh)}-\ref{fi(rh)}) we integrated the system 
(\ref{meq}, \ref{deltaeq}, \ref{weq}, \ref{higseq}) towards $r\to\infty$ using an automatic
step procedure and accuracy $10^{-12}$. The integration stops when the flat spacetime asymptotic
limit (\ref{masimpt}, \ref{deltasimpt}, \ref{wasimpt}, \ref{fiasimpt}) is reached.

The  behaviour of the black-hole  solutions as a function of $\xi$ and $\alpha$
is similar to the regular solutions.  
Two solution branches appear for each $k$ corresponding to two different
values of the shooting parameters $\omega (r_{h})$ and  $\phi(r_{h})$. As
$\alpha\to 0$, the proper $k=1, 2$ branches approaches the corresponding
Einstein-Yang-Mills  black-hole solutions (\cite{6, 7}).

In the same limit, the quasi-$k=0$ branch is distinguished by its Schwarzschild solution limit
 ($\omega =1$, $\phi=0$) and the last node of the  quasi-$k=0$ and  quasi-$k=1$
 branches is again pushed out to infinity. The corresponding limit of the quasi-$k=1$ branch is 
the
 n=1 Einstein-Yang-Mills black-hole solution. 
 
Again, every branch the solutions exist only for a finite range of the parameter 
$0 \leq \alpha \leq \alpha_{max}(\beta,k)$ with different values of  
$\alpha_{max}$ for every solution branch.

The results for $k=$1, 2,  $\beta^{2}$=1/8 and various values of 
the parameter $\xi$ are presented in $figure$ 4.
As we expected, for a nonzero $\xi$ it is necesary to establish new limiting values of the 
values of the normalised vacuum expectation values $\alpha$. For example, a  
minimally coupled solution has necessarily $0<\alpha<0.331$ (proper $k=1$ branch), 
$0<\alpha<0.356$ (quasi-$k=0$ branch), $0<\alpha<0.0475$  (proper $k=2$ branch) and
 $0<\alpha<0.0486$ for the quasi-$k=2$ branch. 

We have found $0<\alpha<0.325$ ($\xi=-1$) and $0<\alpha<0.352$ for $\xi=1/6$ 
(proper $k=1$ branch); some results for the quasi-$k=0$ branch are: $0<\alpha<0.356$ 
($\xi=-0.1$),
$0<\alpha<0.352$ ($\xi=1/6$). For the proper $k=2$ branch we have found  $0<\alpha<0.0382$ 
($\xi=60$),
$0<\alpha<0.336$ ($\xi=-60$), while the quasi-$k=0$ branch with $\xi=10$ has the limiting value
$\alpha<0.044$; for a negative coupling constant $\xi=-5$ we have found $0<\alpha<0.047$ 
(see also $figure$ 5).
Different ranges for the shooting parameters $\omega (r_{h})$ and  $\phi(r_{h})$ are 
to be imposed. 
  
Similar to the case of regular  solutions, we notice the occurence of negative
energy densities.  Anyway, an unexpected  feature is the violation of the WEC even in the
vicinity  of  the  event  horizon (for  positive   values  of $\xi$ and quasi$-k=0$ branch), 
which 
is supposed to destabilize the black-hole and to lead to a traversable wormhole \cite{27}.
  
Another interesting  problem is the effect of the nonminimal coupling 
on the  properties of a black-hole.  Not suprisingly, for the quasi$-k=0$ branch 
we have noticed a violation of the generic relation \cite{27} 
\begin{equation}
T_H = {1\over 4\pi r_H} \; e^{-\delta(r_H)} \; (1-2m'(r_h))
\nonumber\\
\leq\frac{1}{4\pi r_{h}}=T^{vac}_{H}.
\end{equation}
(we use units  $k_{B}=\hbar$=1)
where $T_{H}^{vac}$ is the Hawking temperature of a Schwarzchild black-hole with the same area.

We have found that  generally a positive 
$\xi$ will increase the value of the Hawking temperature (the only disturbing exception is the 
quasi $k=1$ case).                  
However, following \cite{28} the validity of the generic relation $S=\frac{A}{4}$
(where $A$ is the event horizon area) can easily be proven for the Lagrangian 
density (\ref{lag}).


\section{FURTHER DISCUSSION}

Further  insight into the meaning of the nonminimal  coupling in EYMH theory
can be obtained by using the conformal rescaling of the action (\ref{lag}):
\begin{eqnarray} \label{transform1}
\overline g_{\mu\nu}=\Omega^{2} g_{\mu\nu},
\quad
\Omega^{2}=1-2\xi G\phi^{2}.
\end{eqnarray}
The use of this conformal transformation together with a redefinition of the scalar field 
for the case of nonminimal coupling  has 
a long history; ref. \cite{faraoni}, for instance, presents a large set of references on this 
subject.
The usual  condition   $\Omega^{2}>0$ has a clear physical  meaning since it is satisfied by
finite energy solutions only.  The pairs of variables  
(metric $g_{\mu\nu}$, scalar $\phi$, $SU(2)$ field $F_{\mu\nu}$)
defined  originally constitute what is called
a Jordan frame.  Consider now the transformation 
\begin{eqnarray} \label{transform2}
\psi=\int d\phi F(\phi),
\nonumber\\
F^{2}(\phi)=\frac{1-2\xi G\phi^2(1-6\xi)}{(1-2\xi G\phi^{2})^{2}}
\end{eqnarray}
such  that, in the redefined  action
\begin{eqnarray} \label{newaction}
S=\int d^{4}x\sqrt{-\overline g}[\frac{\overline\mathcal{R}}{16\pi G}-
\frac{1}{4\pi}\frac{1}{2}
(\partial_{\mu}\psi)(\partial^{\mu}\psi)
\nonumber\\
-\frac{1}{4\pi}\frac{1}{2}(\frac{g\phi}{2})^{2}
\frac{\mid A\mid ^{2}}{1-2\xi G\phi^{2}}-
\frac{1}{4\pi}\overline V(\psi)
-\frac{1}{4\pi}\frac{1}{4}\mid F\mid^{2}] 
\end{eqnarray}
$\psi$ becomes  minimally  coupled to $\overline\mathcal R$, with 
\begin{equation}
\overline V(\psi)=\frac{V(\phi)}{\Omega^{4}}.
\end{equation}
  
The new variables (metric $\overline g_{\mu\nu}$, scalar $\psi$, 
$SU(2)$ field $F_{\mu\nu}$) are said to constitute an Einstein  frame. The transformation 
given by eqs. (\ref{transform1}, \ref{transform2}) therefore maps a solution of the field 
equations 
imposed by (\ref{lag}) to a solution 
that extremizes (\ref{newaction}). The transformation is independent of any assumption of 
symmetry, and in this sense is covariant; one can easily infer that the 
transformation is one-to-one in general. 
Also, the transformation preserves symmetries, which means that if  $g_{\mu\nu}$ admits a 
Killing vector $\eta$ such that $\pounds_{\eta}\phi=0$ , then $\eta$ is also a Killing vector 
of $\overline g_{\mu\nu}$ and $\pounds_{\eta}\psi=0$.
There is a long debate in the literature on the problem of which of these two  frames  
is  physical (for a review  see \cite{faraoni, magnano}). For example, in ref. \cite{31} 
it has been
shown in a more general context that all thermodynamical variables defined in the
original frame are the same as those in the Einstein frame, if spacetimes in both
frames are asymptotically flat, regular and posses event horizons with non-zero 
temperature. We know that $\Omega^{2}$ goes to some finite positive value at infinity. 
Since 
this value is not unity, the asymptotically Minkowskian metric $g_{\mu\nu}$ will 
be mapped into a generally non-asymptotically Minkowskian line element 
$\overline g_{\mu\nu}$.
However, one needs only to redefine globally the units of length and time 
to obtain an Einstein-frame standard Minkowski form at infinity.  
Considering an expansion of the Higgs field $\phi$ around the minimum 
$\phi=v+\eta$, for large enough negative values of $\xi$ we obtain the following 
first order Einstein frame expression 
\begin{equation}
L_{YM}=
-\frac{1}{4\pi}\frac{1}{4}\mid F\mid^{2}+
\frac{1}{4\pi}\frac{1}{4\xi G}(\frac{g}{4})^{2}
\mid A\mid ^{2}, 
\end{equation}
with an effective decoupling of the YM and Higgs fields and a massive Yang-Mills theory, 
along the line suggested in \cite{24}.

The Weyl rescaling (\ref{transform1}) helps us to rule out the existence of traversable  
wormhole  solutions, since one can conclude that when we know all Einstein-frame solutions with 
a 
given symmetry we automatically know all Jordan-frame solutions with the same symmetry.

 A spacetime  wormhole is usually introduced as a topological handle connecting two universes 
or distant places in the same universe.  Over the last decade following the seminal papers 
of Morris, Thorne and Yurtsever \cite{AmJPhys, Morris},    considerable 
interest has  grown in the domain of traversable wormhole physics (for a
review see \cite{37}).    We recall that a Lorentzian wormhole  solution  is 
said to be traversable  if it does not  contain  horizons that prevent the
crossing of the throat.  A remarkable  result is that,  assuming Einstein 
gravity, the WEC is violated at throat of a traversible static wormhole
\cite{AmJPhys, Hochberg}. 

 Since we have found that the violation of the WEC is possible, it is natural to look for 
spherically  symmetric, traversable wormhole solutions of the coupled EYMH equations. 
Further, it has been conjectured that a violation of the WEC 
in the vicinity of the event  horizon  is quite  likely to  destabilize  the  
horizon  and lead to a traversable  wormhole \cite{27}. Thus, we suppose the  
existence of a traversable  wormhole  solution in the original Jordan frame, 
therefore the violation of the energy condition at or near the throat of the wormhole.
The case of a static spherically symmetric Lorentzian wormhole corresponds to the 
following choice of the metric functions in the general ansatz (\ref{metric})
\begin{eqnarray} 
R(r)=(1-b(r)/r)^{-1/2},
\quad
T(r)=e^{-f(r)},
\end{eqnarray}
where $b(r)$ is called the shape function as it describes the shape of the spatial geometry of 
the wormhole in an embedding diagram and $f(r)$ describes the gravitational redshift in this 
spacetime (with $e^{f(r)}>0$) \cite{AmJPhys, 37}. 
In this case the coordinate $r$ is constrained to run between $r_{0}<r<\infty$, where $r_{0}$
 is the throat radius ($b(r_{0})=r_{0}$). 
By following the analysis of Sec. 2, we can again predict the general features
 of the possible solutions and boundary conditions. 
It follows that the general relations obtained as $r\to\infty$ are still valid. 
If we do not allow for $\phi$ or $\phi '$ to take an infinite value at the
 wormhole throat, a similar analysis of the Higgs field equation (\ref{fi2}) 
implies that the generic conditions 
$\Omega^{2}>0$,  $\xi<\frac{1}{2\alpha^{2}}$ hold also . 
  
Using the conformal transformation (\ref{transform1}) we convert the theory to the Einstein  
frame. The existence of wormhole solutions is  not  affected  by  the transformation 
(\ref{transform1}) 
that preserves the traversability for $\Omega^{2}>0$, i.e. a 
positive effective gravitational constant and no event horizon in the new frame.
It can easily be proven  that for the  rescaled  action (\ref{newaction}), the  dominant
energy condition holds, and thus there are no traversable  wormhole  solutions, i.e. 
no traversable wormhole solutions in the Jordan frame also.  

One can wonder whether this absence of traversable wormhole solutions
is a general feature of arbitrary nonminimal scalar couplings to Einstein
gravity. The nonminimal coupling of the scalar field
considered in  this paper is a particular case of a more general theory,
where the term    $\frac{\mathcal{R}}{16\pi G}(1-2 \xi G\phi^{2})$
is  replaced by a more general function $\frac{\mathcal{R}}{16\pi G}f(\phi)$.

By using a conformal transformation  
$\overline g_{\mu\nu}= f(\phi) g_{\mu\nu}$
and a redefinition of the scalar field (\cite{faraoni})
\begin{eqnarray}
\psi=\int d\phi (\frac{f(\phi)+\frac{3}{4G} (\frac{df(\phi)}{d\phi})^2}{f(\phi)^2})^{1/2}
\end{eqnarray}
we can convert the action to the Einstein frame
\begin{eqnarray} 
S=\int d^{4}x\sqrt{-\overline g}[\frac{\overline\mathcal{R}}{16\pi G}-
\frac{1}{4\pi}\frac{1}{2}
(\partial_{\mu}\psi)(\partial^{\mu}\psi)
\nonumber\\
-\frac{1}{4\pi}\frac{1}{2}(\frac{g\phi}{2})^{2}
\frac{\mid A\mid ^{2}}{f(\phi)}-
\frac{1}{4\pi}\overline V(\psi)
-\frac{1}{4\pi}\frac{1}{4}\mid F\mid^{2}] 
\end{eqnarray}
(with $\overline V(\psi)=\frac{V(\phi)}{f(\phi)^{2}}$).

If $f(\phi)>0$ (i.e. a positive effective Newton constant) the WEC will be satisfied in 
the Einstein frame. The positivity of the effective Newton constant followed
in our case from the demand that the energy of the solution is finite; implying
$\xi<\frac{1}{2\alpha^{2}}$.  We have not been able to show that an equivalent
condition holds in the general case.  The possibility of traversable wormholes
in general is therefore left open, though it seems likely they will be absent.

\section{CONCLUSIONS}

In this paper we have studied static, spherically symmetric classical solutions
of spontaneously broken SU(2) gauge theory with a nonminimally coupled Higgs field
and presented strong numerical arguments for the existence of both regular and 
black-hole solutions for suitable values of the coupling parameter $\xi$.  
The main properties of these solutions, such as their nodal 
structure and discrete mass spectrum, are generic and shared by
practically all known solitons with gravitating non-abelian gauge fields.

It should be stressed that it is the non-abelian nature of the Yang-Mills field that
 allows 
the existence of nontrivial solutions since a nonminimally coupled 
Higgs charged hair has been ruled out for Abelian Higgs theory 
(\cite{Adler}, see also \cite {Mayo}). 

The absence of 
 gravitating sphalerons and sphaleron black-holes in a spontaneously broken 
theory of gravity has also been proven. 
As a new feature we have established  a violation of the WEC for a certain range of $\xi$.
The nonexistence of traversable wormhole 
solutions has been shown using a conformal map to convert the problem  to
the one with minimal coupling to gravity.
For small values of the parameter $\xi$,  the effect of the nonminimal coupling 
on the asymptotic 
features of a finite energy solution
is rather benign. Although we did not address the nature of the solutions for 
$r<r_{h}$, we expect  that a term $\xi\Phi^{2}\mathcal{R}$ can strongly influence
the properties of the inner black-hole solutions. 

	We have not  considered  the question of stability of solutions  in 
this paper.
Since in the case of minimal coupling  the solutions  were found to be  unstable,  
we see no  reason  to expect something special to happen for  $\xi\ne0$.\\

{\bf Acknowledgement}
\newline
 This work was performed in the context of the
Graduiertenkolleg of the Deutsche Forschungsgemeinschaft (DFG):
Nichtlineare Differentialgleichungen: Modellierung,Theorie, Numerik, Visualisierung.



\newpage
Figure Captions
\newline

Figure 1:
The parameter beta versus the maximal value of the parameter $\alpha$ for 
one node graviting sphaleron solutions; qualitative picture for the minimally coupled case.  
\newline

Figure 2:
One- and two-node sphaleron solutions of the nonminimally coupled EYMH theory
 for $\alpha=0.1$, $\beta^{2}=1/8$ and various values of $\xi$.
\newline

Figure 3:
One- and two-node sphaleron solutions of the nonminimally coupled EYMH theory
for $\beta^{2}=1/8$ and various values of $\xi$.
Here and in $figure$ 5, the parameter $\alpha$ varies between zero and 
the maximum allowed value $\alpha_{max}$; increasing $\alpha$ corresponds to 
a decrease of the value of the radial coordinate at which the solution 
exponentially decays to its vacuum value.
\newline

Figure 4:
One- and two-node black hole solutions of the nonminimally coupled EYMH theory
for $\beta^{2}=1/8$ and various values of $\xi$.
\newline

Figure 5:
One- and two-node black hole solutions of the nonminimally coupled EYMH theory
for $\beta^{2}=1/8$ and various values of $\xi$.

\newpage
\setlength{\unitlength}{1cm}

\begin{picture}(16,16)
\centering
\put(-2,0){\epsfig{file=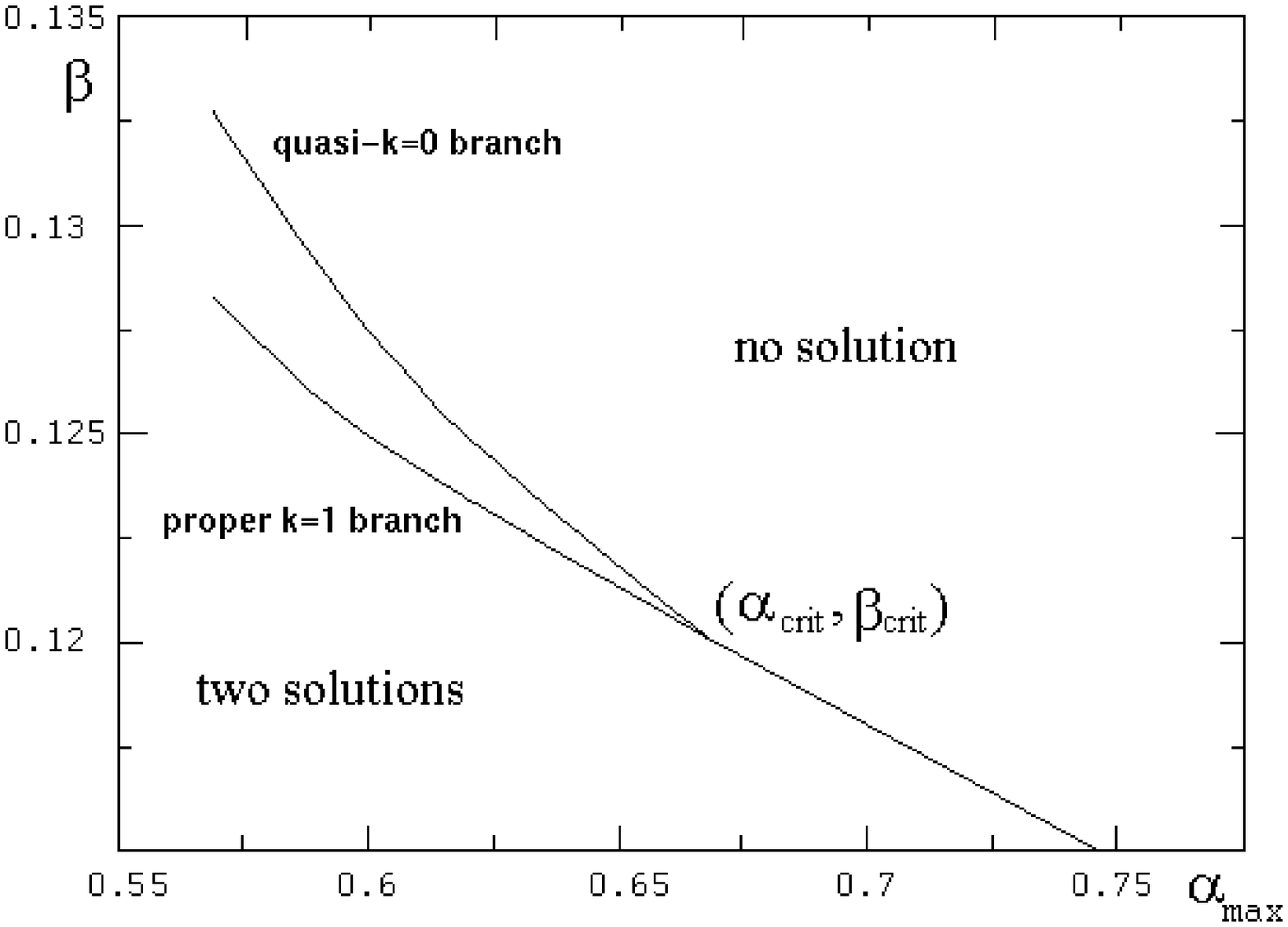,width=18cm}}
\end{picture}
\begin{center}
Figure 1.
\end{center}

\newpage
\setlength{\unitlength}{1cm}

\begin{picture}(16,16)
\centering
\put(-2,0){\epsfig{file=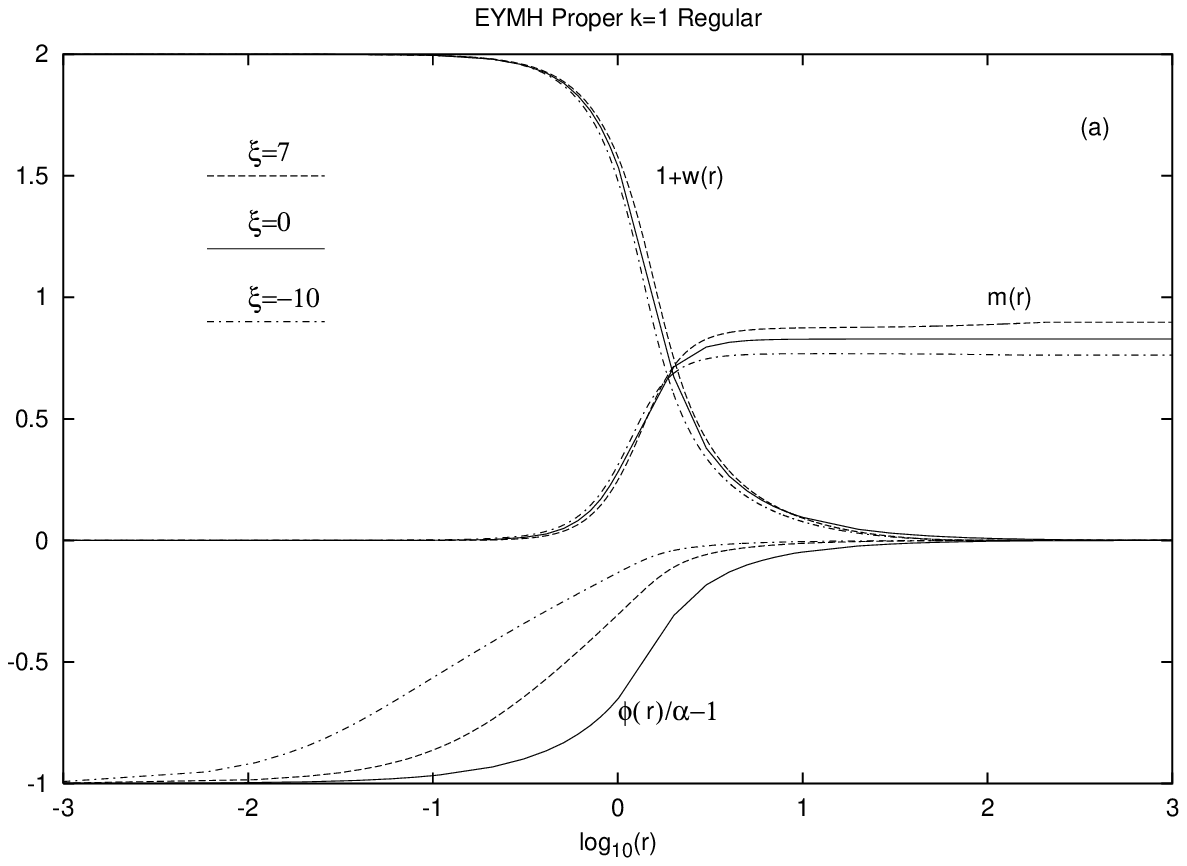,width=16cm}}
\end{picture}
\begin{center}
Figure 2a.
Proper $k=1$ Regular $\xi=7$, 0, -10\newline
\end{center}

\newpage
\begin{picture}(16,16)
\centering
\put(-2,0){\epsfig{file=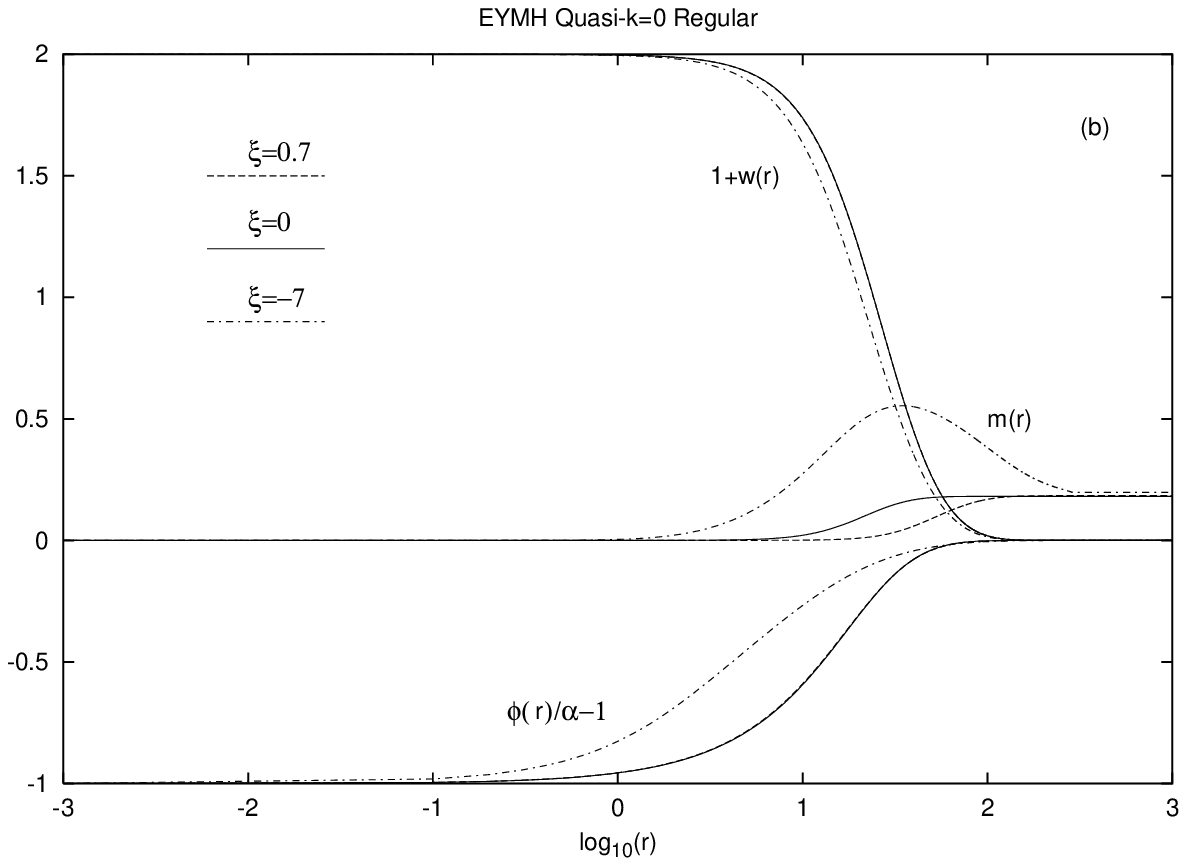,width=16cm}}
\end{picture}
\begin{center}
Figure 2b.
Quasi$-k=0$ Regular $\xi=0.7$, 0, -7\newline
\end{center}

\newpage
\begin{picture}(16,16)
\centering
\put(-2,0){\epsfig{file=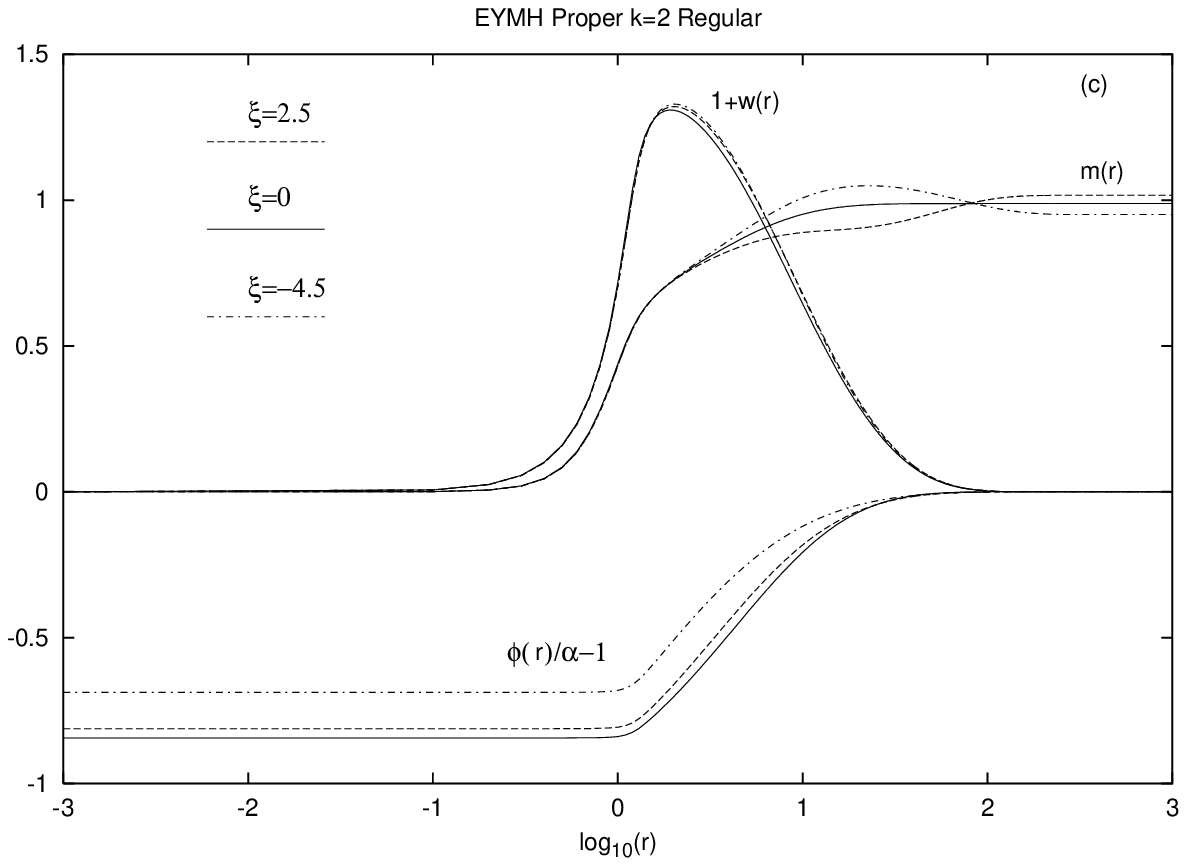,width=16cm}}
\end{picture}
\begin{center}
Figure 2c.
Proper $k=2$ Regular $\xi=2.5$, 0, -4.5\newline
\end{center}

\newpage
\begin{picture}(16,16)
\centering
\put(-2,0){\epsfig{file=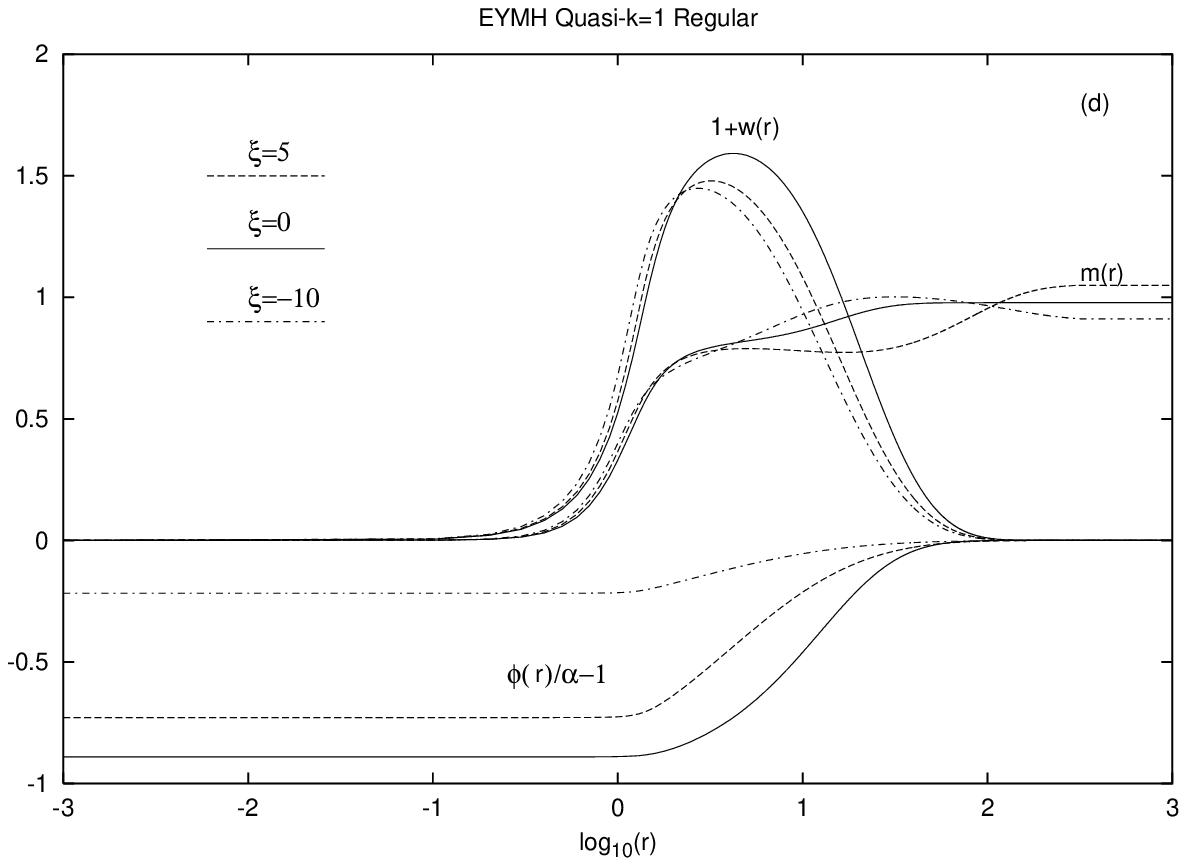,width=16cm}}
\end{picture}
\begin{center}
Figure 2d.
Quasi$-k=1$ Regular $\xi=5$, 0, -10\newline
\end{center}

\newpage
\begin{picture}(16,16)
\centering
\put(-2,0){\epsfig{file=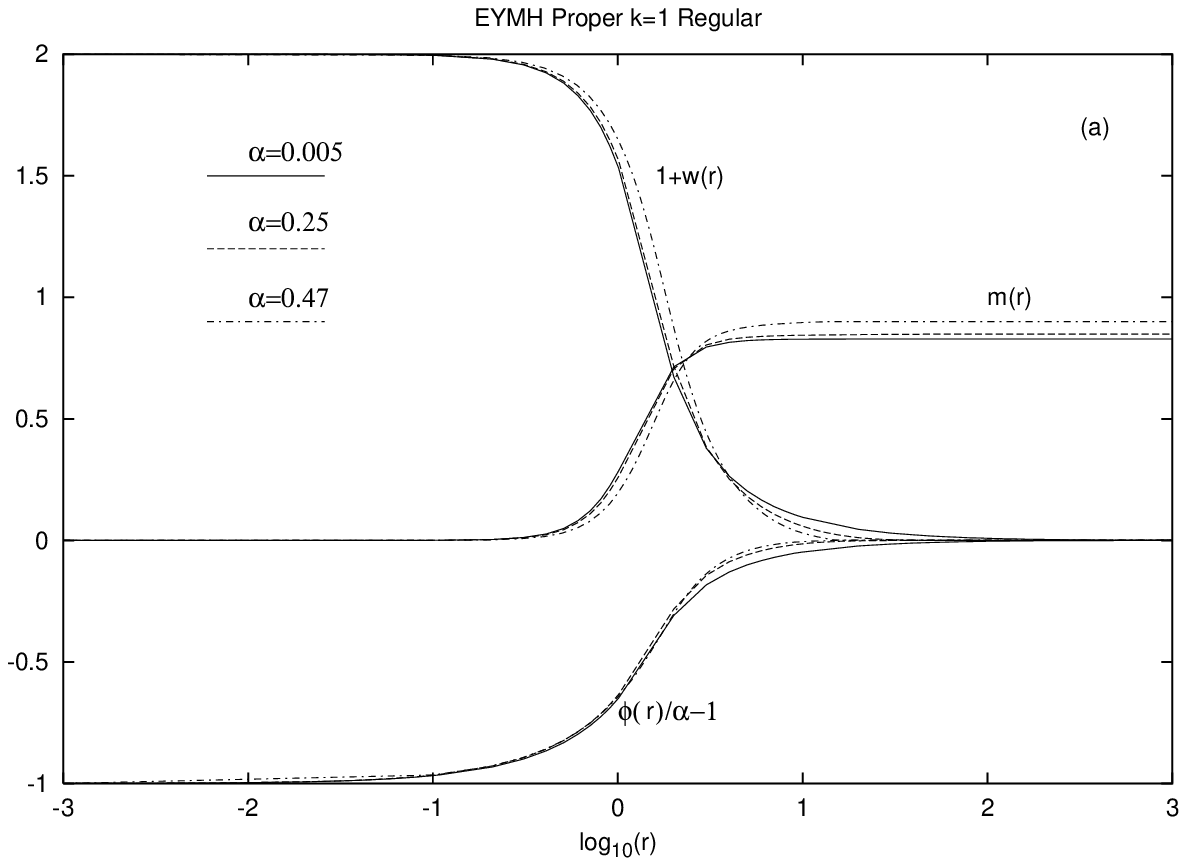,width=16cm}}
\end{picture}
\begin{center}
Figure 3a.
Proper $k=1$ Regular $\xi=0.1$; { }$\alpha=0.005$, 0.25, 0.47\newline
\end{center}

\newpage
\begin{picture}(16,16)
\centering
\put(-2,0){\epsfig{file=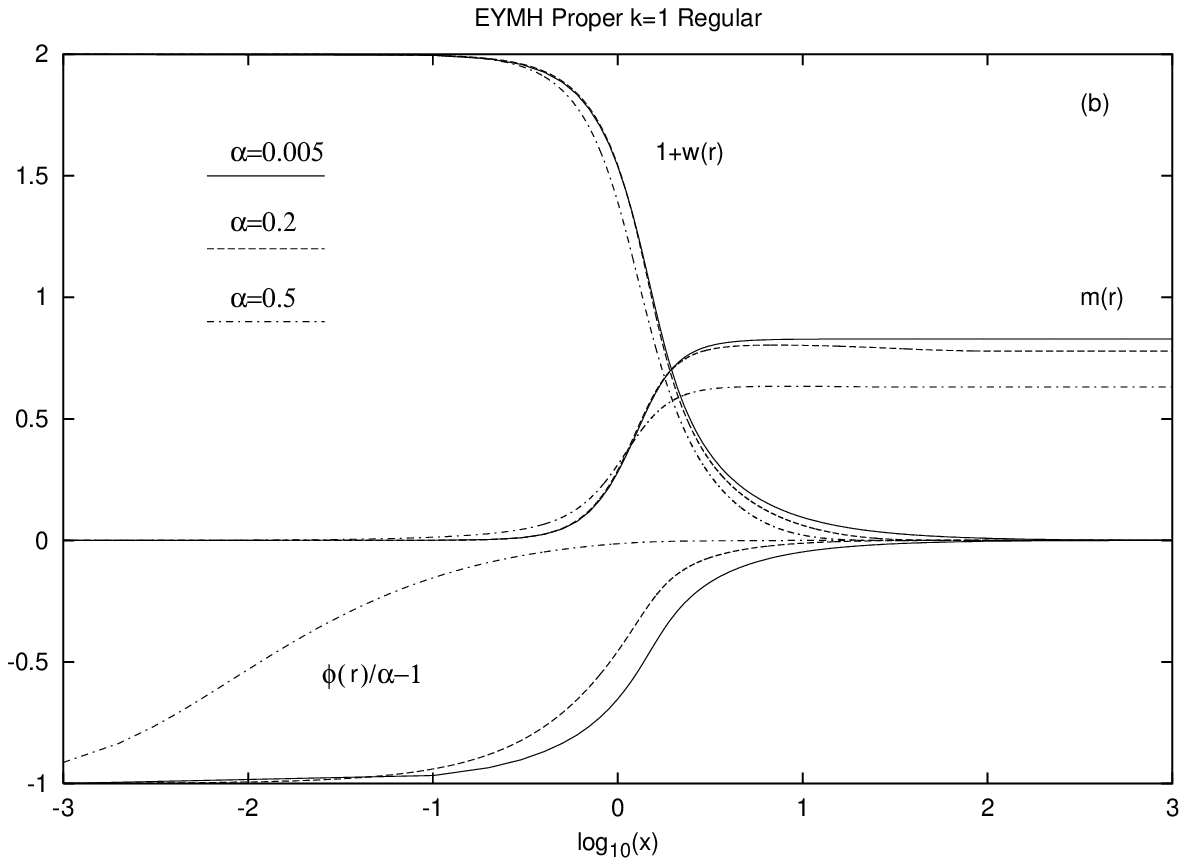,width=16cm}}
\end{picture}
\begin{center}
Figure 3b.
Proper $k=1$ Regular $\xi=-2$; { }$\alpha=0.005$, 0.2, 0.5\newline
\end{center}

\newpage
\begin{picture}(16,16)
\centering
\put(-2,0){\epsfig{file=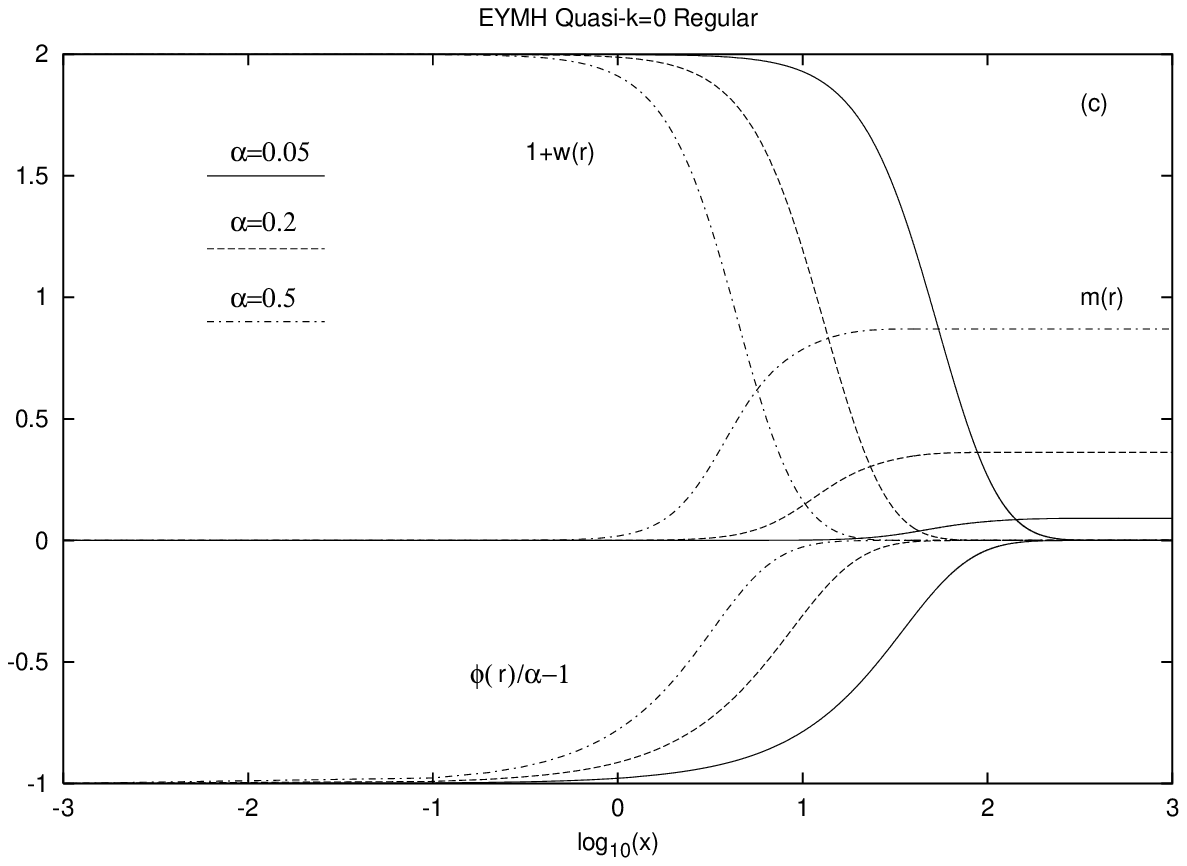,width=16cm}}
\end{picture}
\begin{center}
Figure 3c.
Quasi$-k=0$ Regular $\xi=1/6$; { }$\alpha=0.005$, 0.2, 0.5\newline
\end{center}

\newpage
\begin{picture}(16,16)
\centering
\put(-2,0){\epsfig{file=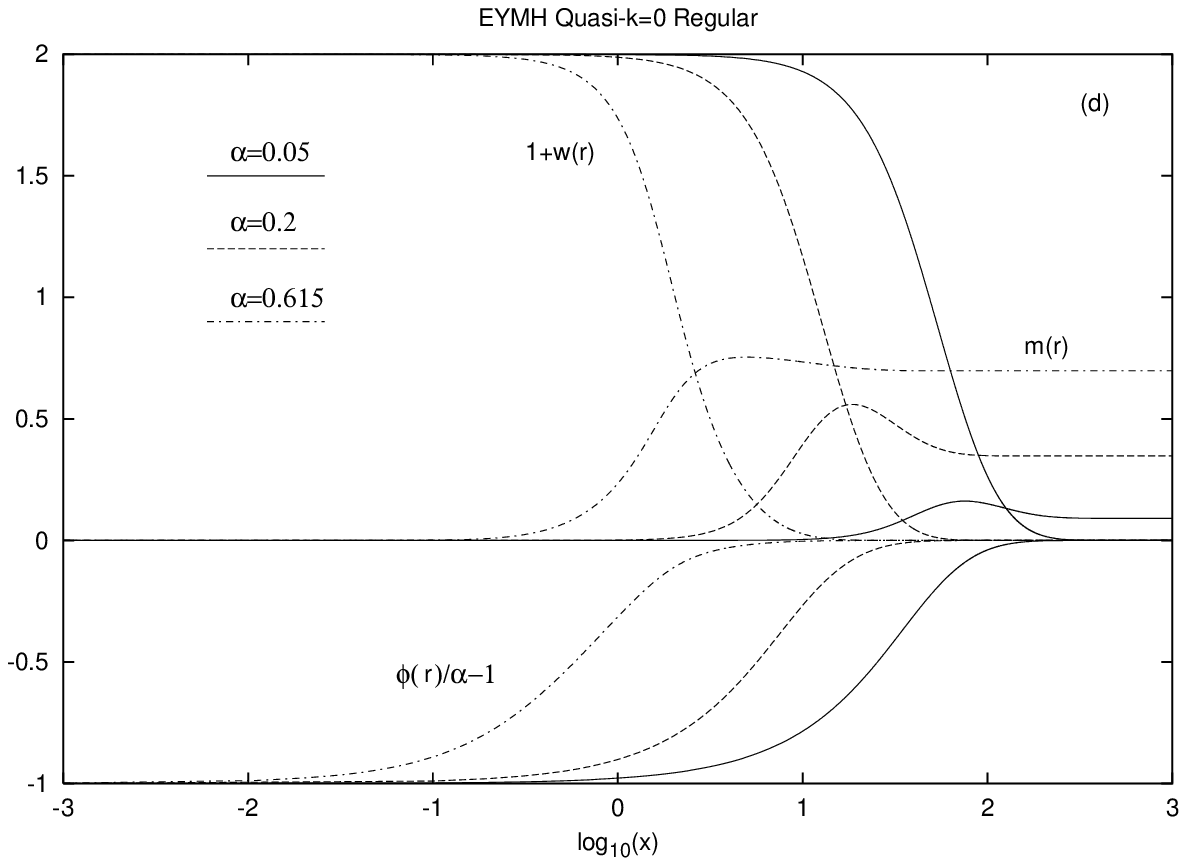,width=16cm}}
\end{picture}
\begin{center}
Figure 3d.
Quasi$-k=0$ Regular $\xi=-1$; { }$\alpha=0.05$, 0.2, 0.615\newline
\end{center}

\newpage
\begin{picture}(16,16)
\centering
\put(-2,0){\epsfig{file=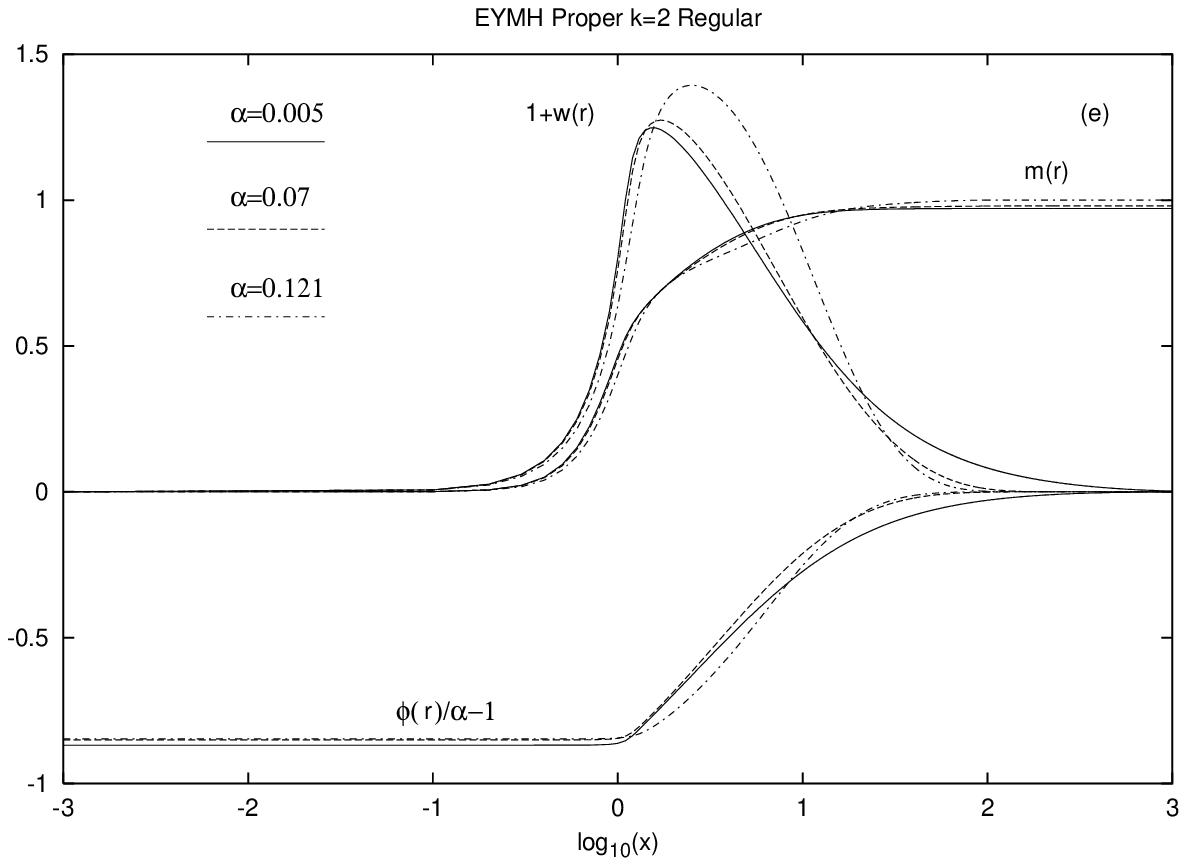,width=16cm}}
\end{picture}
\begin{center}
Figure 3e.
Proper $k=2$ Regular $\xi=1/6$; { }$\alpha=0.005$, 0.07, 0.121\newline
\end{center}

\newpage
\begin{picture}(16,16)
\centering
\put(-2,0){\epsfig{file=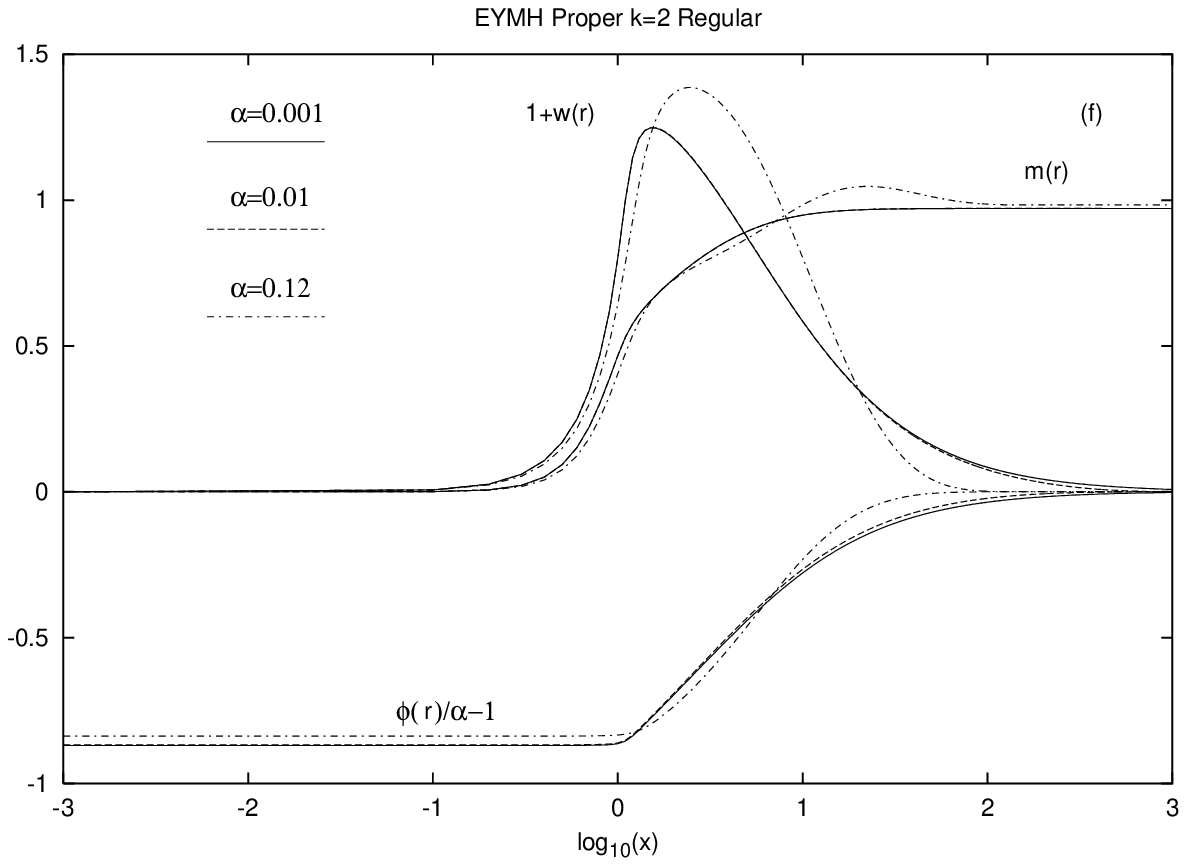,width=16cm}}
\end{picture}
\begin{center}
Figure 3f.
Proper $k=2$ Regular $\xi=-1$;{ }$\alpha=0.001$, 0.01, 0.12\newline
\end{center}

\newpage
\begin{picture}(16,16)
\centering
\put(-2,0){\epsfig{file=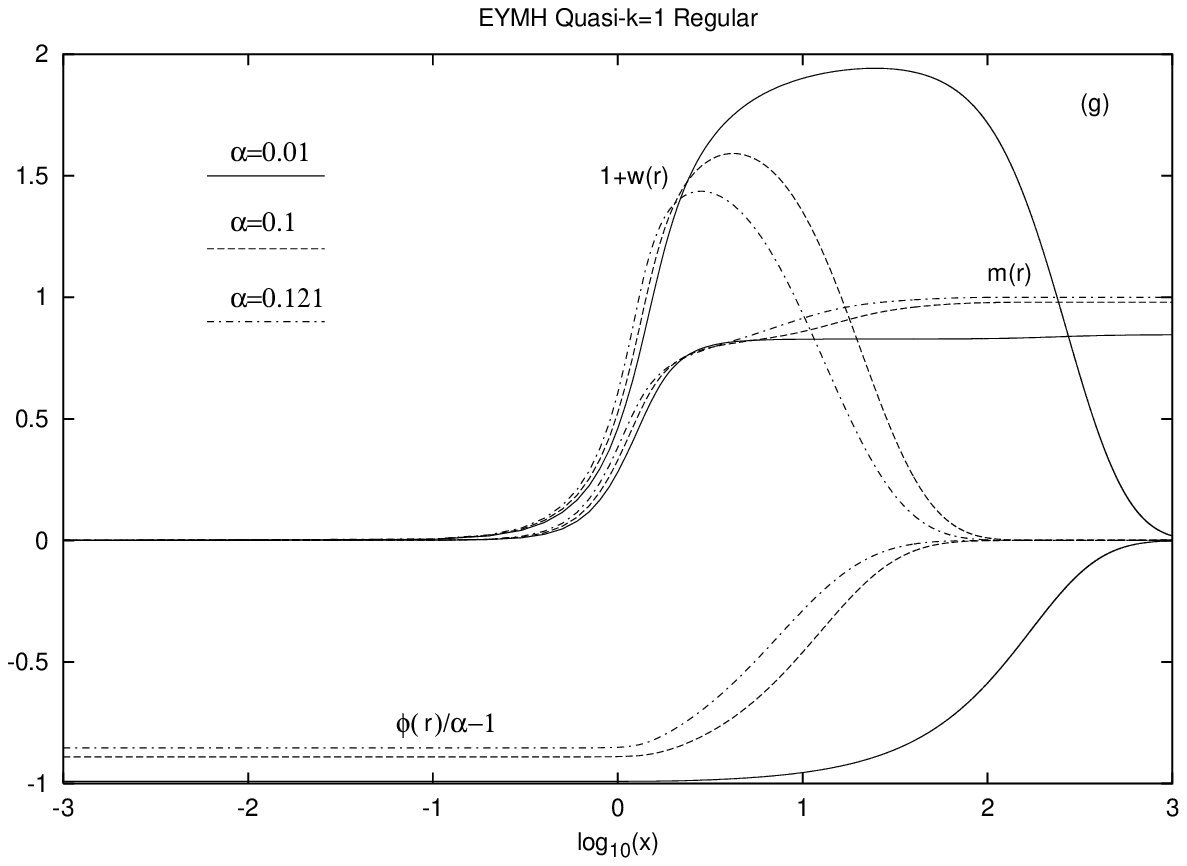,width=16cm}}
\end{picture}
\begin{center}
Figure 3g.
Quasi$-k=1$ Regular $\xi=1/6$; { }$\alpha=0.01$, 0.1, 0.121\newline
\end{center}

\newpage
\begin{picture}(16,16)
\centering
\put(-2,0){\epsfig{file=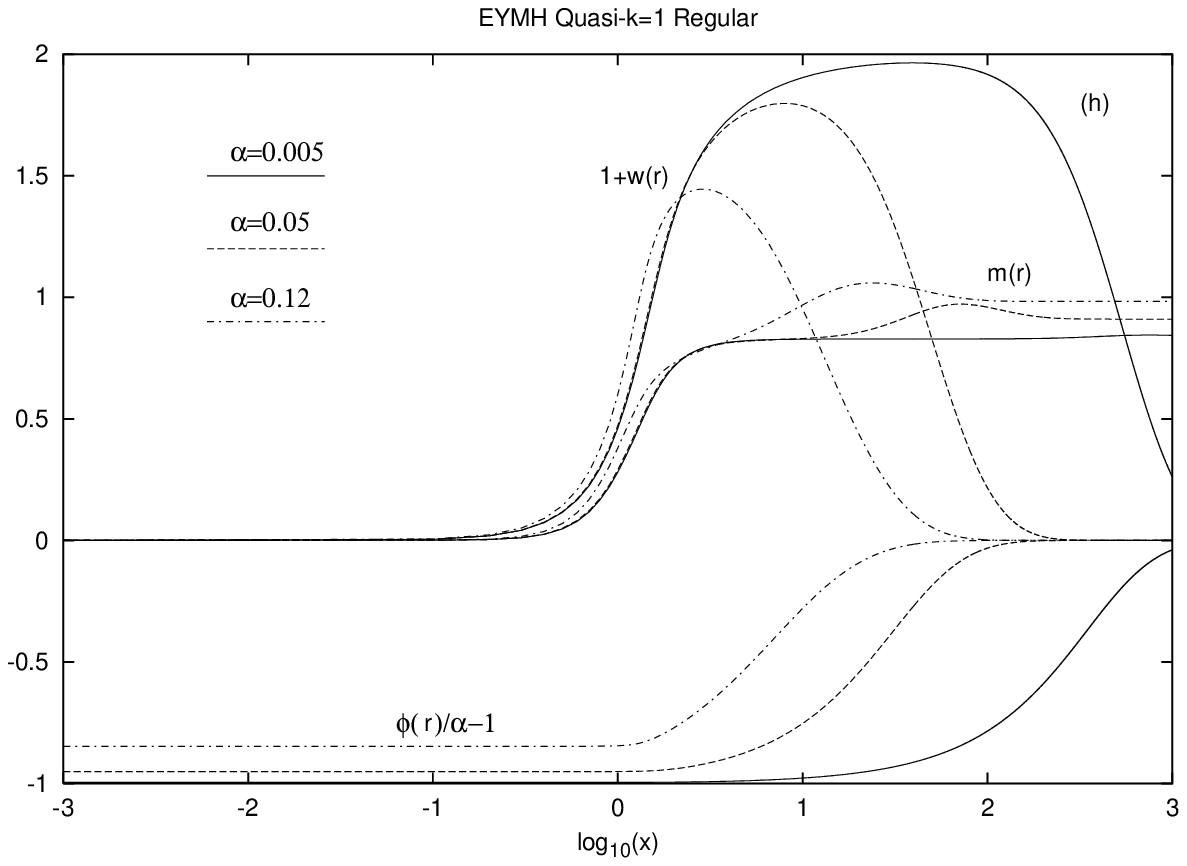,width=16cm}}
\end{picture}
\begin{center}
Figure 3h.
Quasi$-k=1$ Regular $\xi=-1$; { }$\alpha=0.005$, 0.05, 0.12\newline
\end{center}

\newpage
\begin{picture}(16,16)
\centering
\put(-2,0){\epsfig{file=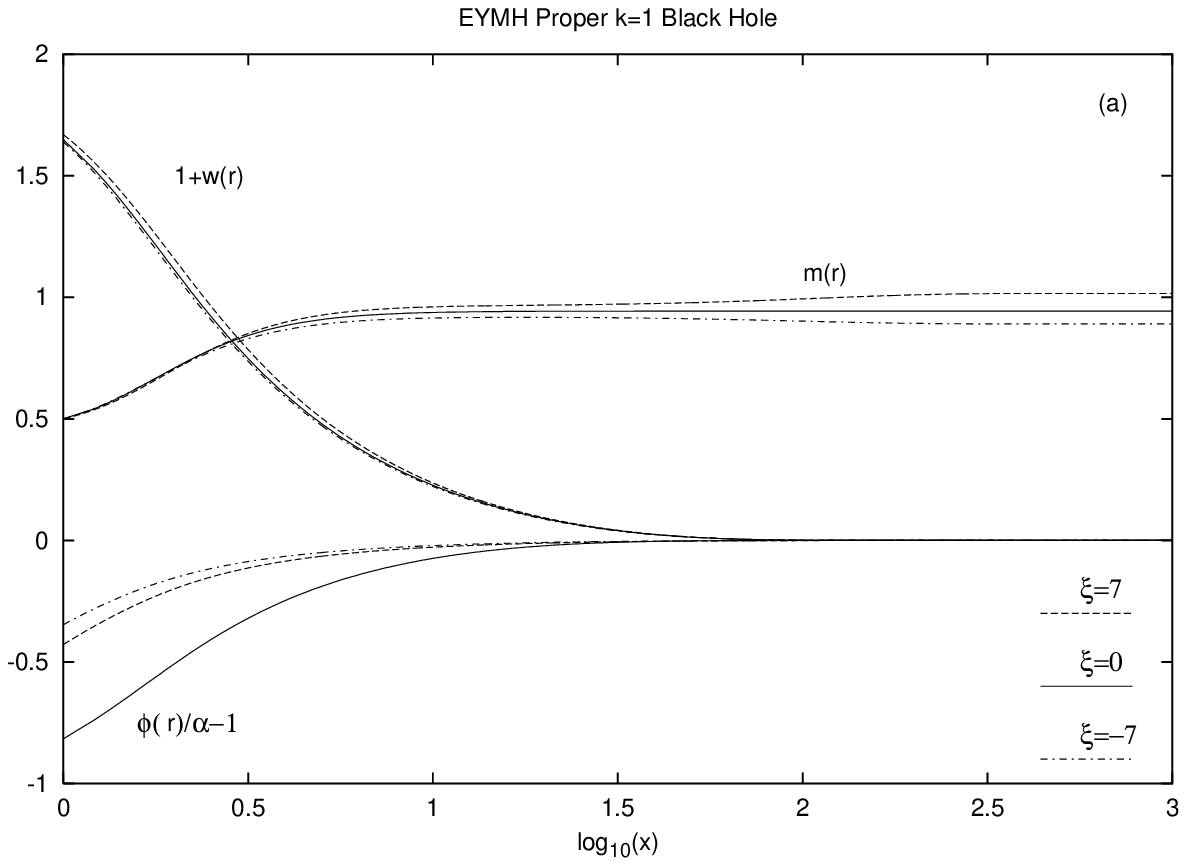,width=16cm}}
\end{picture}
\begin{center}
Figure 4a.
Proper $k=1$ Black Hole; { }$\alpha=0.1$; $\xi=7$, 0, -7\newline
\end{center}

\newpage
\begin{picture}(16,16)
\centering
\put(-2,0){\epsfig{file=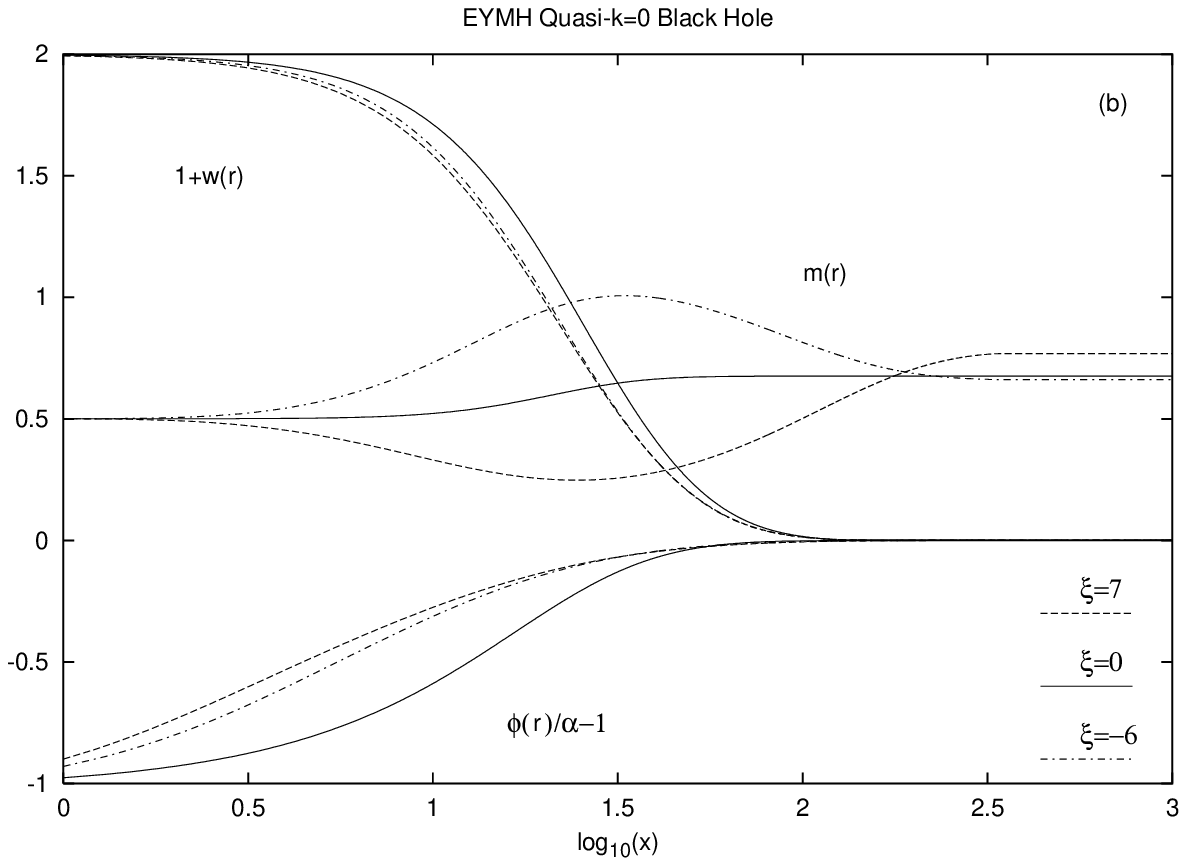,width=16cm}}
\end{picture}
\begin{center}
Figure 4b.
Quasi$-k=0$ Black Hole; { }$\alpha=0.1$;  $\xi=7$, 0, -6\newline
\end{center}

\newpage
\begin{picture}(16,16)
\centering
\put(-2,0){\epsfig{file=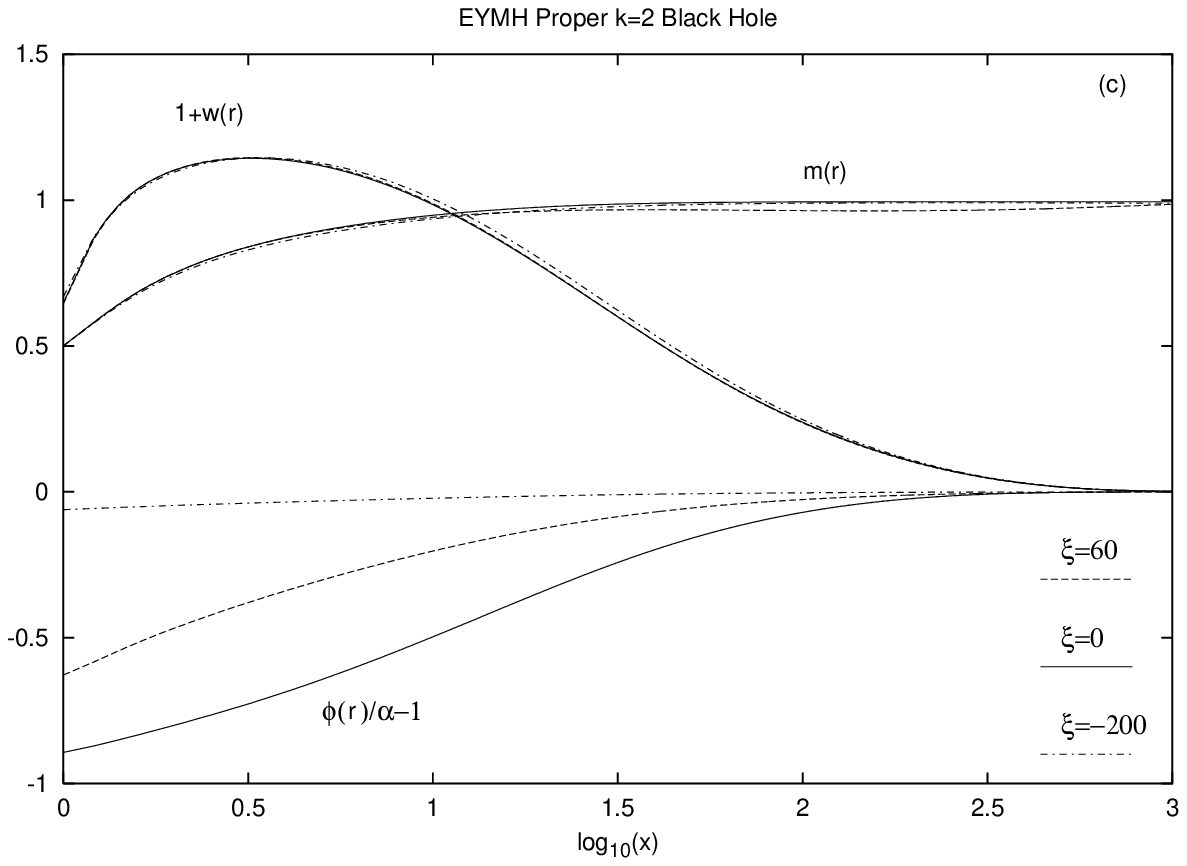,width=16cm}}
\end{picture}
\begin{center}
Figure 4c.
Proper $k=2$ Black Hole; { }$\alpha=0.01$;  $\xi=60$, 0, -200\newline
\end{center}

\newpage
\begin{picture}(16,16)
\centering
\put(-2,0){\epsfig{file=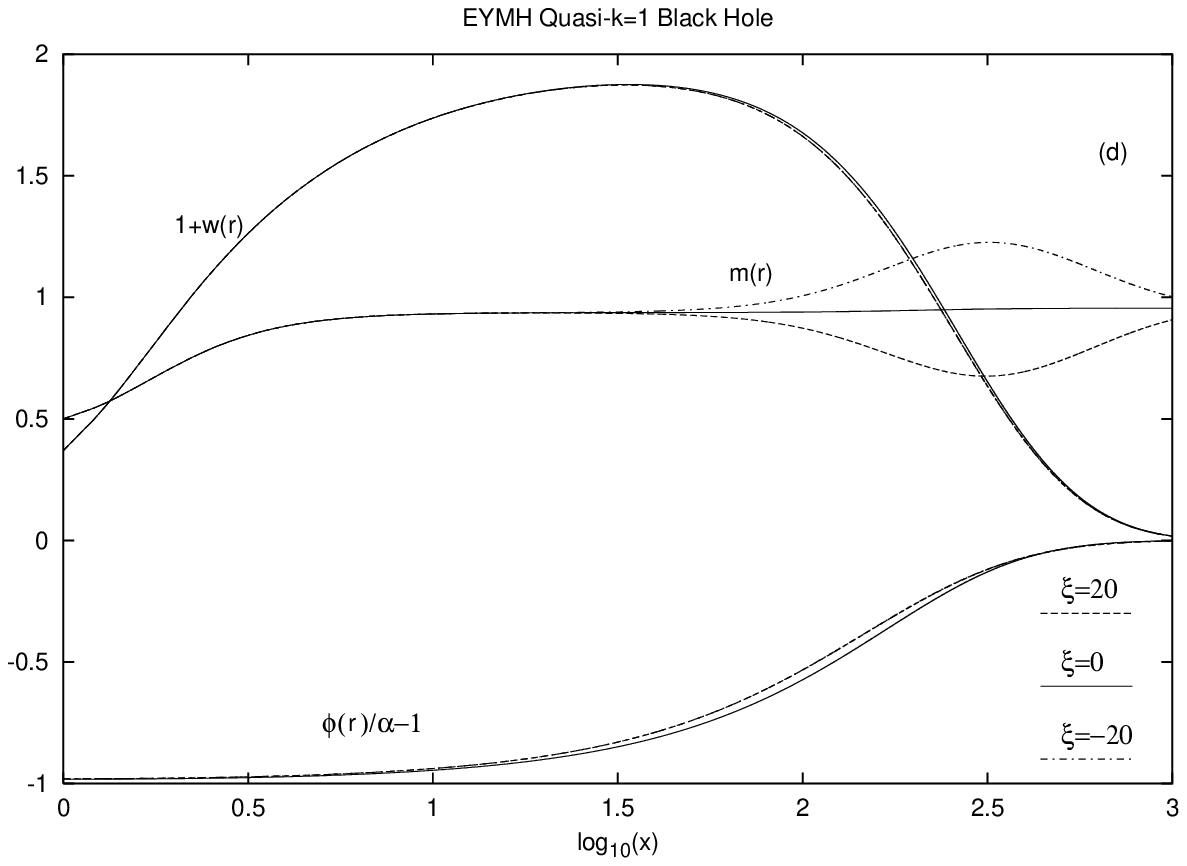,width=16cm}}
\end{picture}
\begin{center}
Figure 4d.
Quasi$-k=1$ Black Hole; { }$\alpha=0.01$;  $\xi=20$, 0, -20\newline
\end{center}

\newpage
\begin{picture}(16,16)
\centering
\put(-2,0){\epsfig{file=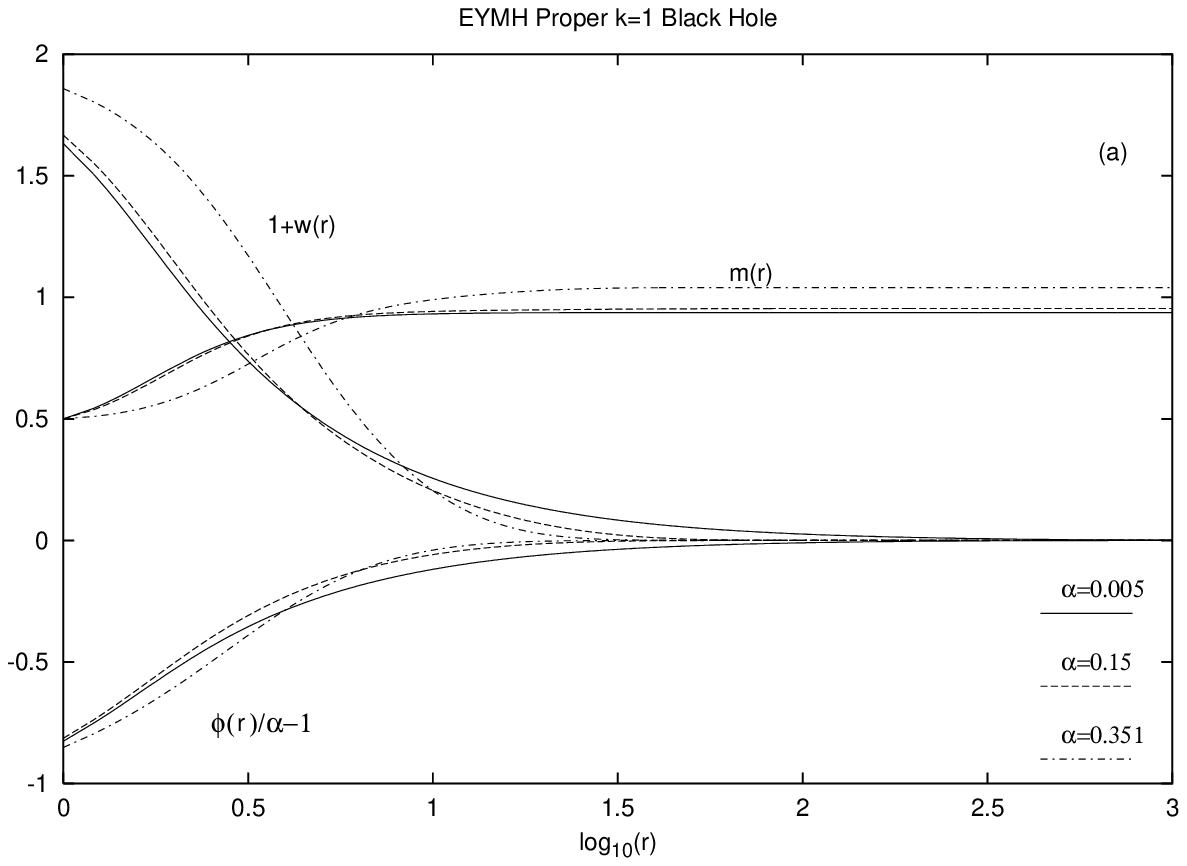,width=16cm}}
\end{picture}
\begin{center}
Figure 5a.
Proper $k=1$ Black Hole $\xi=1/6$; { }$\alpha=0.005$, 0.15, 0.351\newline
\end{center}

\newpage
\begin{picture}(16,16)
\centering
\put(-2,0){\epsfig{file=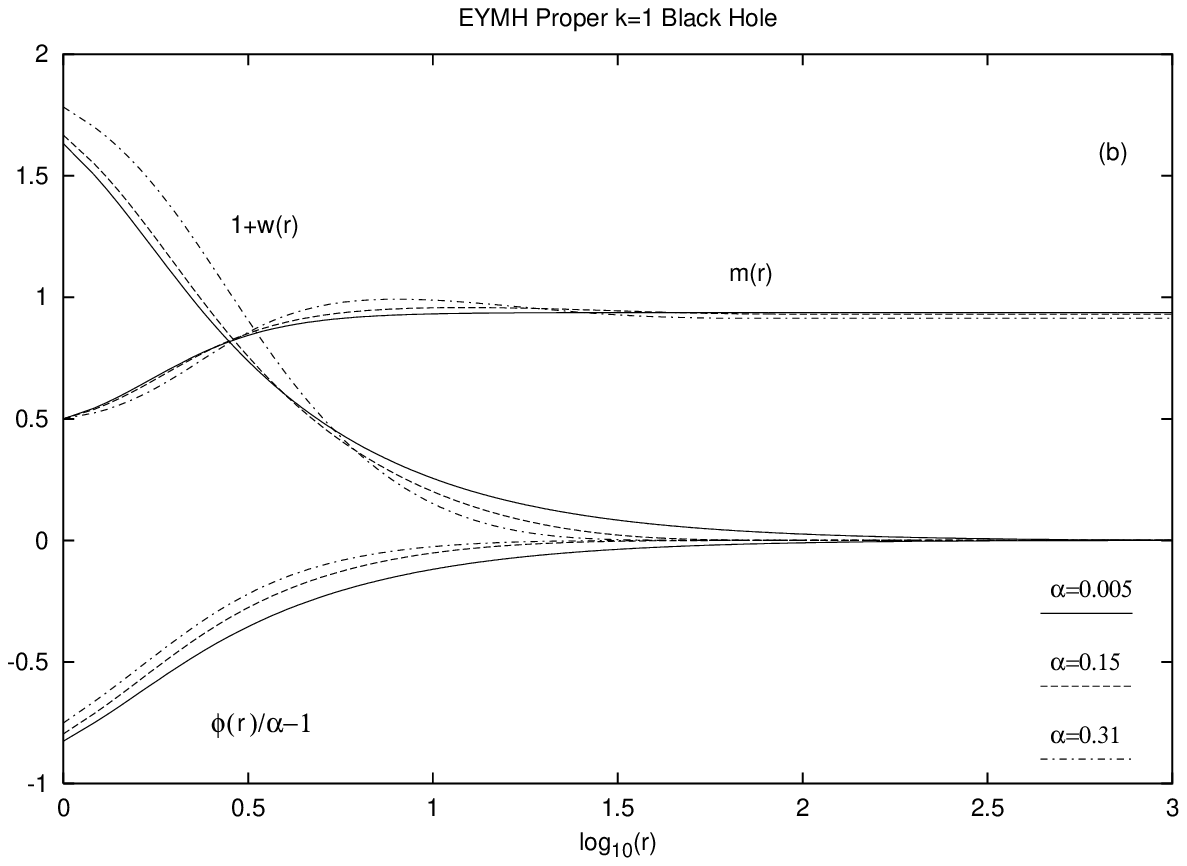,width=16cm}}
\end{picture}
\begin{center}
Figure 5b.
Proper $k=1$ Black Hole $\xi=-1$; { }$\alpha=0.005$, 0.15, 0.31\newline
\end{center}

\newpage
\begin{picture}(16,16)
\centering
\put(-2,0){\epsfig{file=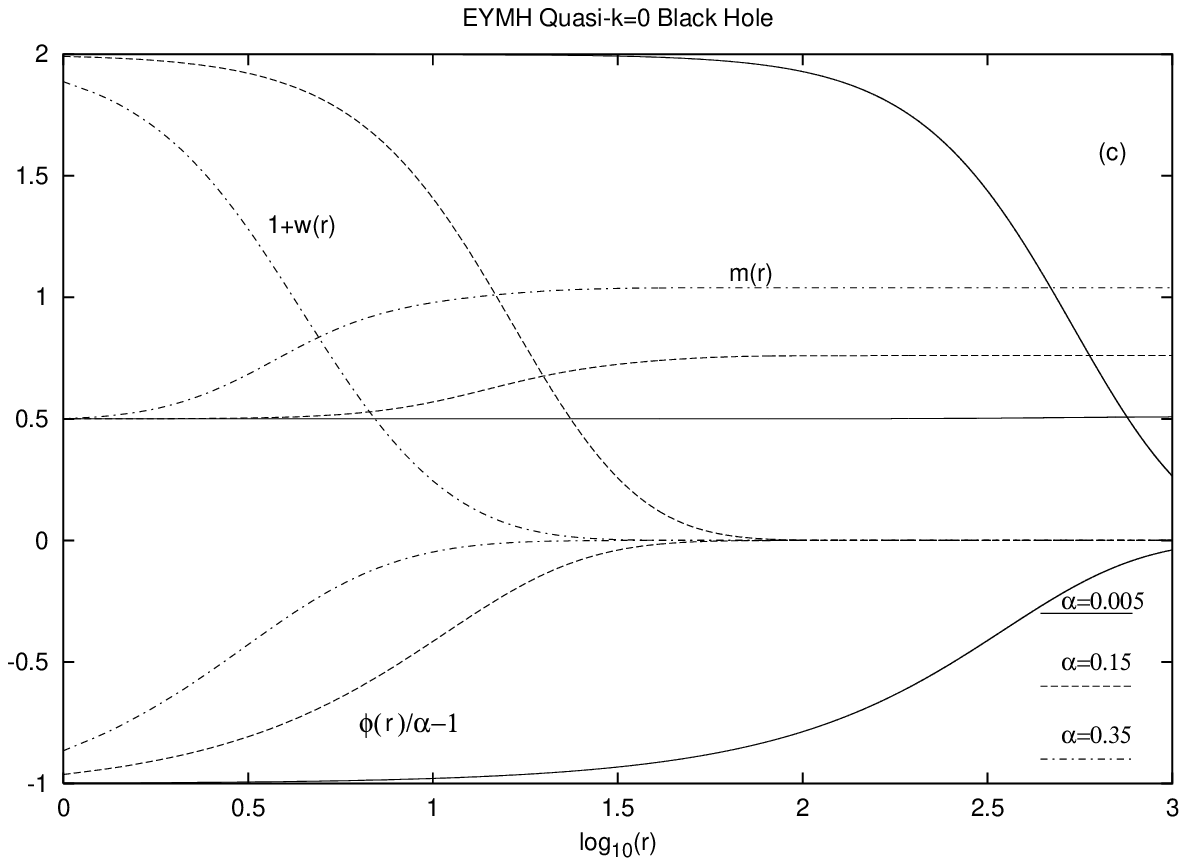,width=16cm}}
\end{picture}
\begin{center}
Figure 5c.
Quasi$-k=0$ Black Hole $\xi=1/6$; { }$\alpha=0.005$, 0.15, 0.35\newline
\end{center}

\newpage
\begin{picture}(16,16)
\centering
\put(-2,0){\epsfig{file=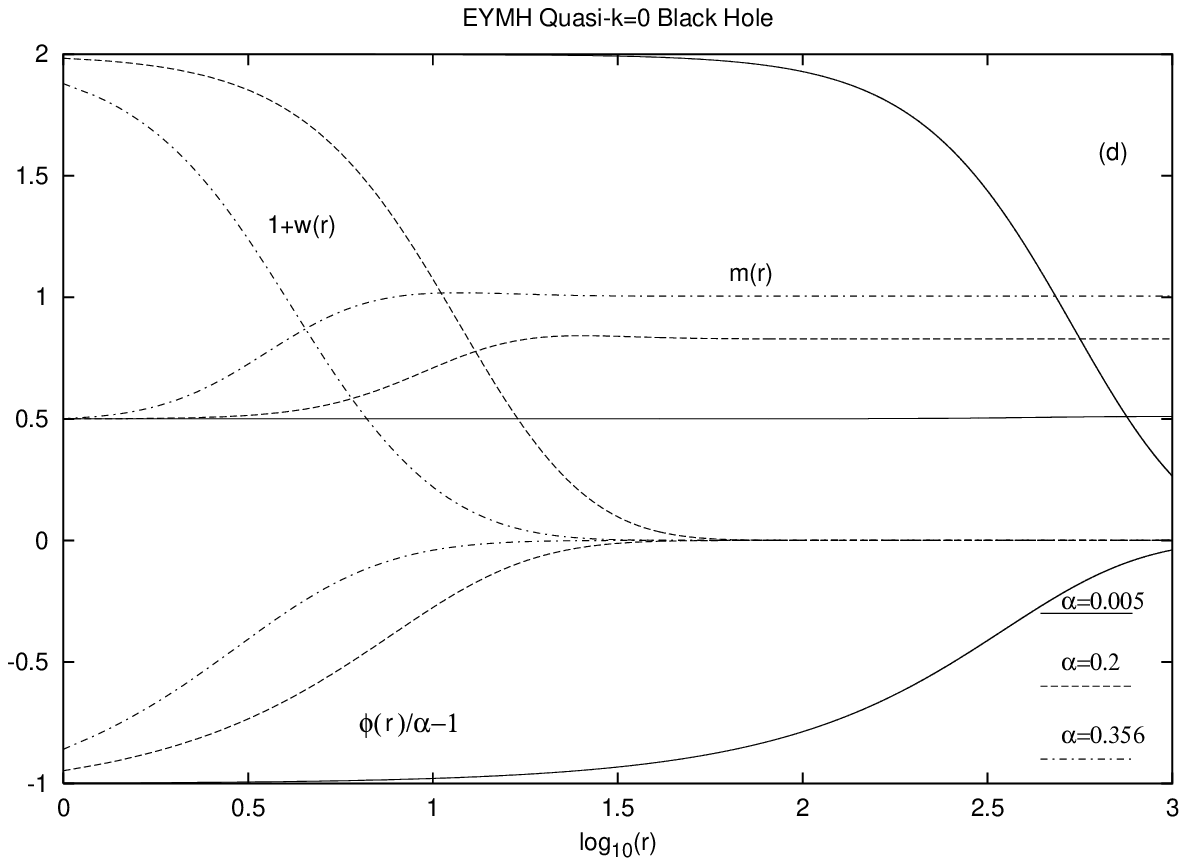,width=16cm}}
\end{picture}
\begin{center}
Figure 5d.
Quasi$-k=0$ Black Hole $\xi=-0.1$; { }$\alpha=0.005$, 0.2, 0.356\newline
\end{center}

\newpage
\begin{picture}(16,16)
\centering
\put(-2,0){\epsfig{file=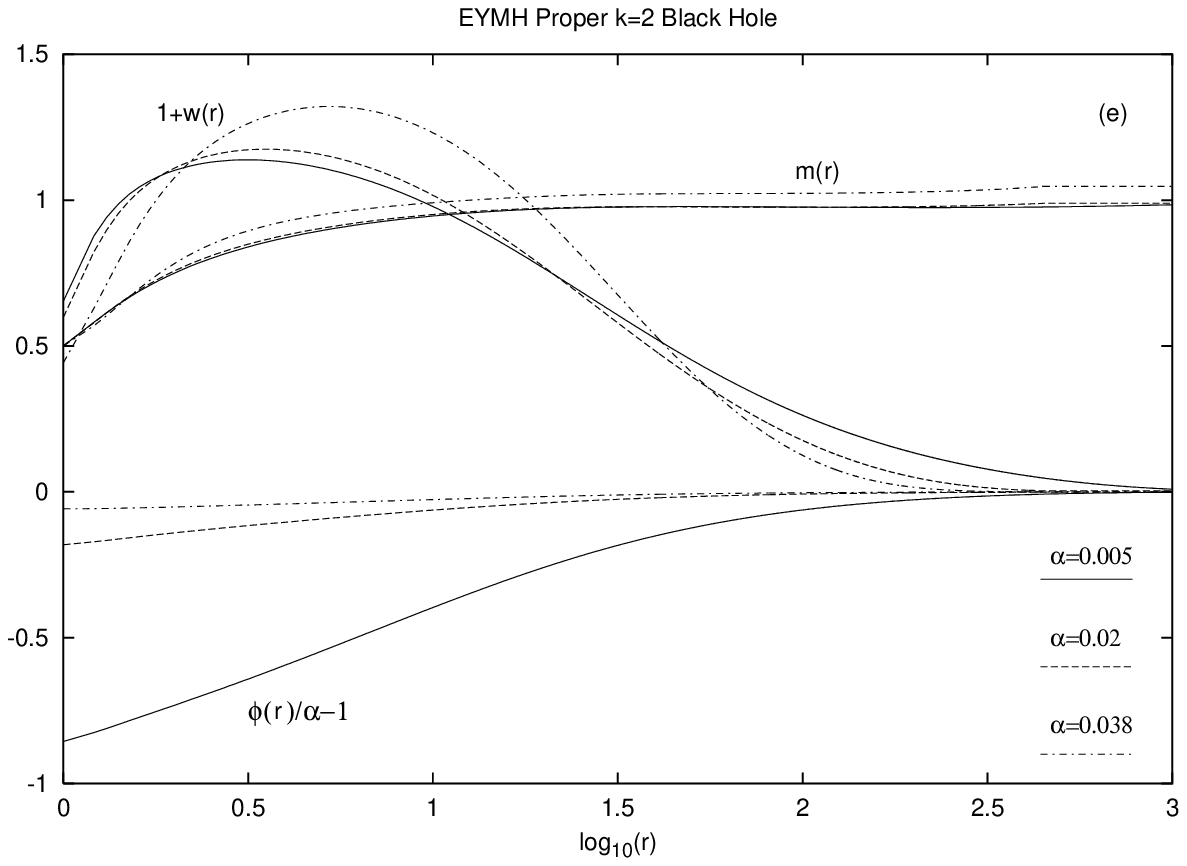,width=16cm}}
\end{picture}
\begin{center}
Figure 5e.
Proper $k=2$ Black Hole $\xi=60$; { }$\alpha=0.005$, 0.02, 0.038\newline
\end{center}

\newpage
\begin{picture}(16,16)
\centering
\put(-2,0){\epsfig{file=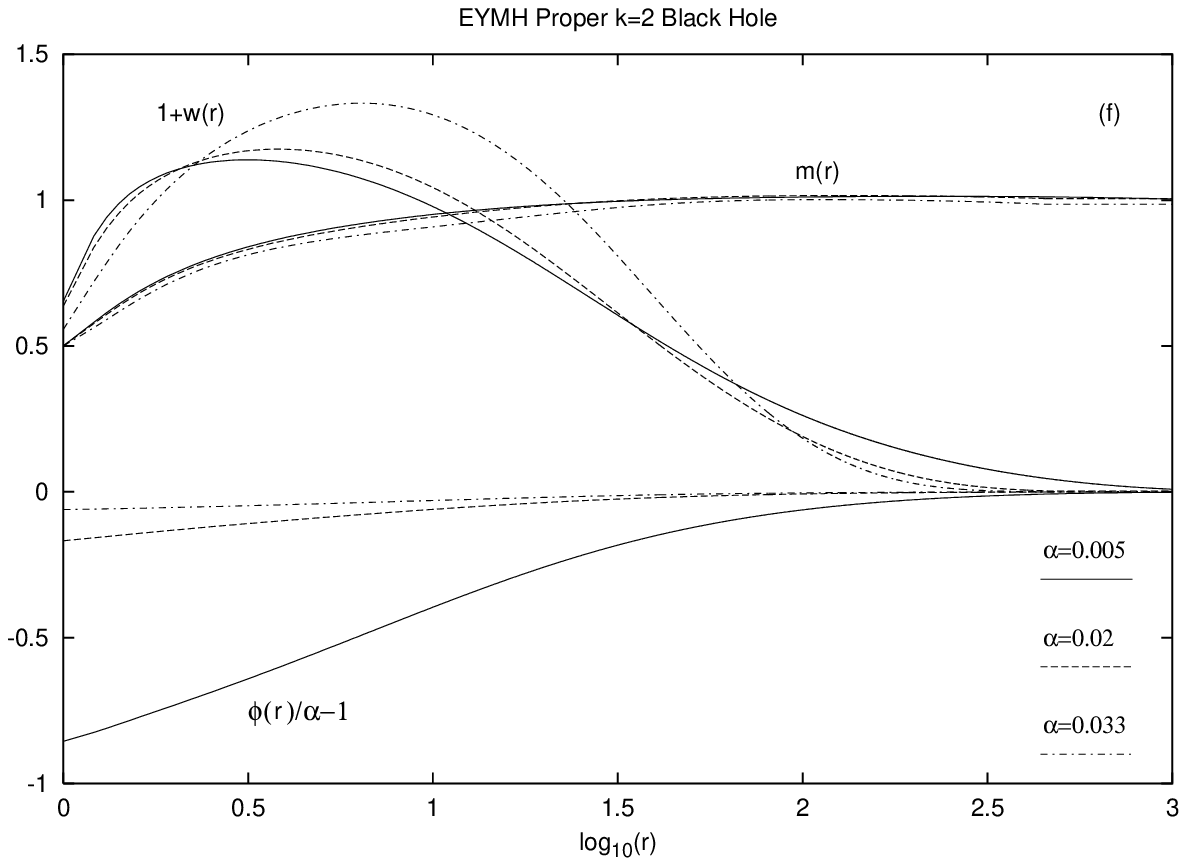,width=16cm}}
\end{picture}
\begin{center}
Figure 5f.
Proper $k=2$ Black Hole $\xi=-60$; { }$\alpha=0.005$, 0.02, 0.033\newline
\end{center}

\newpage
\begin{picture}(16,16)
\centering
\put(-2,0){\epsfig{file=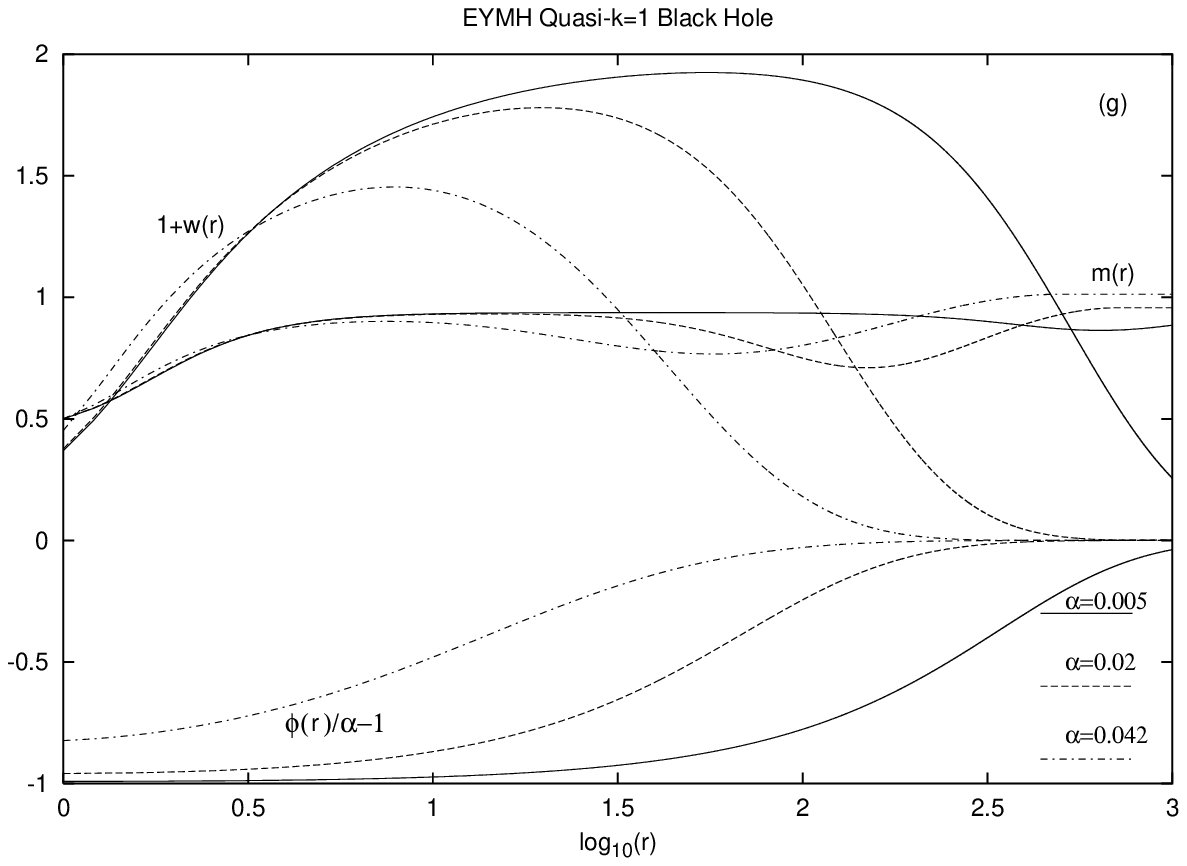,width=16cm}}
\end{picture}
\begin{center}
Figure 5g.
Quasi-$k=1$ Black Hole $\xi=10$; { }$\alpha=0.005$, 0.02, 0.042\newline
\end{center}

\newpage
\begin{picture}(16,16)
\centering
\put(-2,0){\epsfig{file=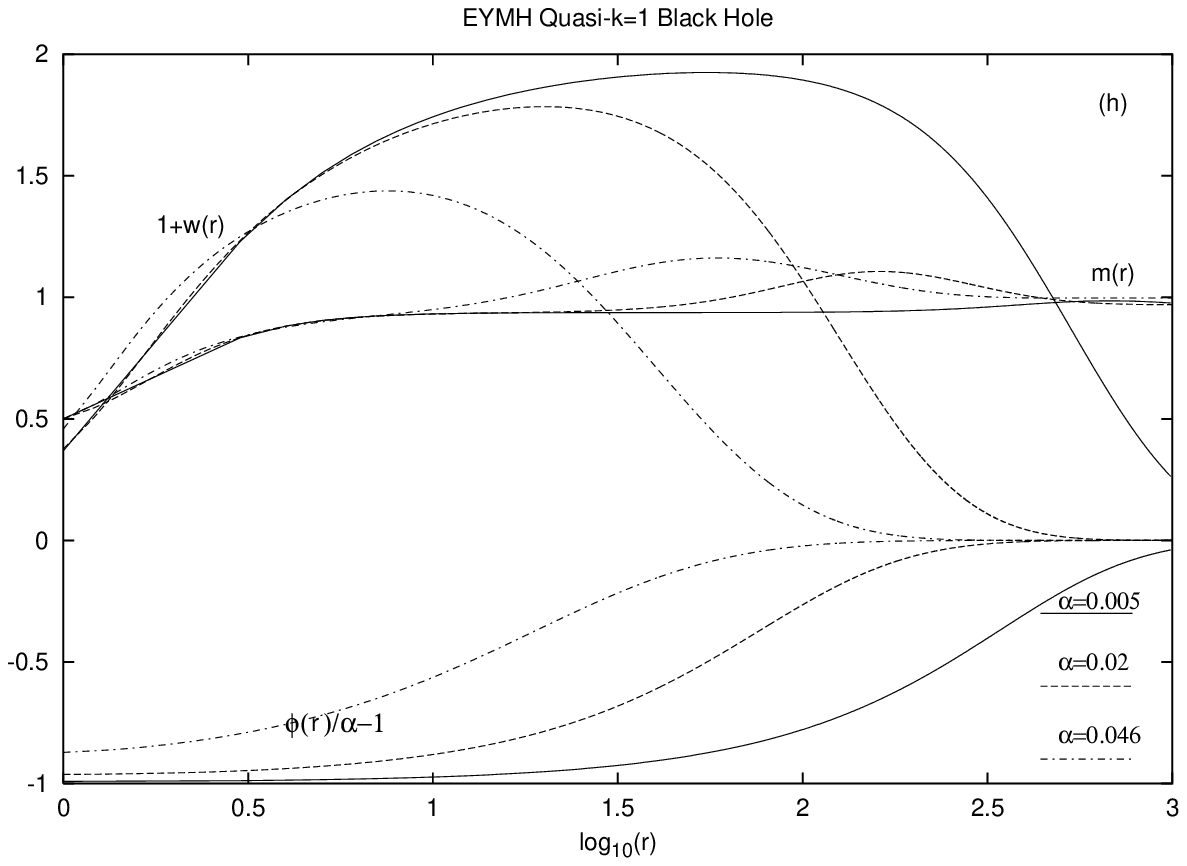,width=16cm}}
\end{picture}
\begin{center}
Figure 5h.
Quasi-$k=1$ Black Hole $\xi=-5$; { }$\alpha=0.005$, 0.02, 0.046\newline
\end{center}

\end{document}